\documentclass{ws-ijmpa}
\usepackage[super,compress]{cite}
\usepackage{graphicx}
\usepackage{subfigure}
\usepackage{lineno}
\newcommand{\avg}[1]    {\ensuremath{\langle #1 \rangle}\xspace} % for e.g. <\mu>

\newcommand{\Herwig}    {\texttt{Herwig}\xspace}

\newcommand{\geant}     {\texttt{GEANT}4\xspace}

\newcommand{\Madgraph}  {\texttt{MadGraph}\xspace}
\newcommand{\Pythia}    {\texttt{PYTHIA}\xspace}
\newcommand{\Powheg}    {\texttt{POWHEG}\xspace}

\newcommand{\PythiaP}   {\Pythia (\Perugia 2010)\xspace}

\newcommand{\antikt}{anti-$k_{t}$}
\newcommand{\Antikt}{Anti-$k_{t}$}

\newcommand{\pt}        {p_T}

\newcommand{\pttrue}  {\ensuremath{\pt^\mathrm{true}}}

\newcommand{\ptreco}  {\ensuremath{\pt^{\mathrm{reco}}}}

\newcommand{\jetarea}     {\ensuremath{A^{\mathrm{jet}}}\xspace}

\newcommand{\avgmu}   {\ensuremath{\avg{ \mu }}\xspace}

\newcommand{\Z}         {\Zboson\xspace}

\newcommand{\Zjets}     {\Zboson+jets\xspace}

\input{boostedsymbols.tex}

\def \et {E_{T}}
\def  \met {\not\!\!\et }
\def \gj {$\gamma$+jet}

\def \Herwig {\texttt{Herwig}\xspace}
\def \Pythia  {\texttt{PYTHIA}\xspace}
\def \PythiaS  {\texttt{PYTHIA6}\xspace}
\def \PythiaP  {\texttt{PYTHIA8}\xspace}
\def \HerwigPP  {\texttt{Herwig}++\xspace}
\def \geant     {\texttt{GEANT}4\xspace}
\def \Madgraph {\texttt{MadGraph}\xspace}

\def \Z    {\ensuremath{Z}}
\def \Zjets  {$Z$+jets}
\def \Zb  {$Z$+b}

\begin{document}
\markboth{Ariel Schwartzman}
{Instructions for Typing Manuscripts (Paper's Title)}

\catchline{}{}{}{}{}
%
%%%%%%%%%%%%%%%%%%%%%%%%%%%%%%%%%%%%%%%%%%%%%%%%%%%%%%%%%%%%%%%%%%%%

\title{JET ENERGY CALIBRATION AT THE LHC}

\author{ARIEL SCHWARTZMAN}

\address{SLAC National Accelerator Laboratory, 2575 Sand Hill Road\\
Menlo Park, California 94025,
USA\\
sch@slac.stanford.edu}

\maketitle

\begin{history}
\received{Day Month Year}
\revised{Day Month Year}
\end{history}

\begin{abstract}
Jets are one of the most prominent physics signatures of high energy proton proton (p-p) collisions at the Large Hadron Collider (LHC). They are key physics objects for precision measurements and searches for new phenomena. This review provides an overview of the reconstruction and calibration of jets at the LHC during its first Run. ATLAS and CMS developed different approaches for the reconstruction of jets, but use similar methods for the energy calibration. ATLAS reconstructs jets utilizing input signals from their calorimeters and use charged particle tracks to refine their energy measurement and suppress the effects of multiple p-p interactions ({\it pileup}). CMS, instead, combines calorimeter and tracking information to build jets from {\it particle flow} objects. Jets are calibrated using Monte Carlo (MC) simulations and a residual {\it in situ} calibration derived from collision data is applied to correct for the differences in jet response between data and Monte Carlo. Large samples of dijet, \Zjets, and \gj~ events at the LHC allowed the calibration of jets with high precision, leading to very small systematic uncertainties. Both ATLAS and CMS achieved a jet energy calibration uncertainty of about 1\% in the central detector region and for jets with transverse momentum $p_T>100$ GeV. At low jet $p_T$, the jet energy calibration uncertainty is less than 4\%, with dominant contributions from pileup, differences in energy scale between quark and gluon jets, and jet flavor composition.

\end{abstract}

\ccode{PACS numbers:}

%\tableofcontents

%-----------------------------------------------------------
\section{Introduction}	
%-----------------------------------------------------------

\begin{figure}[h]
  \centering
    \includegraphics[width=12.0cm]{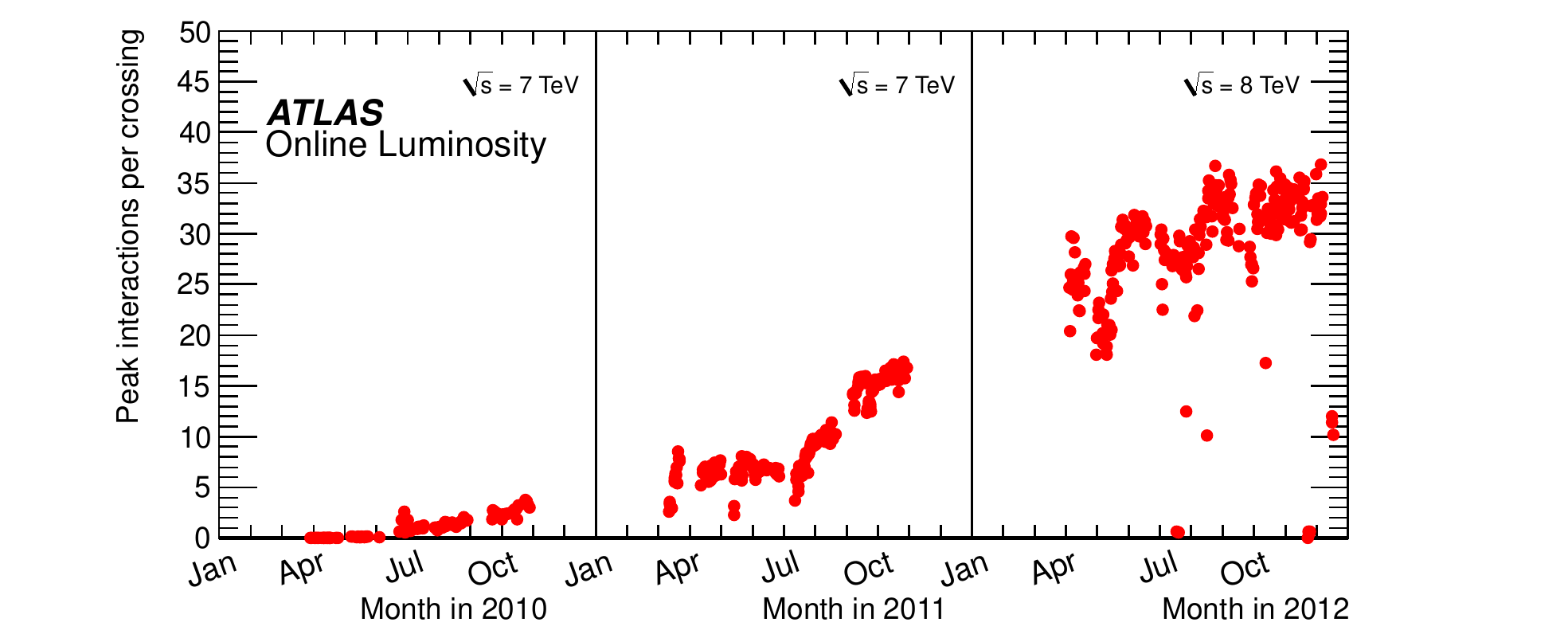}
\caption{Evolution of the maximum mean number of events per beam crossing during the p-p runs of 2010, 2011, and 2012. Figure taken from Ref. \protect\refcite{lumi}.}
\label{fig:pileup}
\end{figure}

One set of key observables in any event at the LHC are collimated streams of hadrons known as {\it jets}. Jets are the observable manifestation of quarks and gluons and are {\it defined} by particle clustering {\it algorithms}. Individual jets are proxies for quark and gluons. Combinations of jets are used to identify unstable massive particles such as the top quark, and the W, Z, and Higgs bosons. Much of the success of the LHC physics program rests on the ability to reconstruct jets and measure their energy accurately. One of the most difficult challenges for the reconstruction and calibration of jets at the LHC is the presence of pileup: additional p-p interactions produced within the same event. Pileup reduces the accuracy of the jet energy measurement and can produce additional jets that do not originate from the hard-scatter interaction. Pileup conditions at the LHC changed significantly through Run 1, as shown in Fig. \ref{fig:pileup}. Initially, during 2010, the low proton beam intensities resulted in a very small amount of pileup. The increase in proton beam intensity in 2011 and 2012 led to a corresponding increase in the average number of pileup interactions per bunch crossing from an average of 3 in 2010 to approximately 8 and 20 during the 2011 and 2012 data taking periods respectively. Despite this challenging environment, both LHC experiments have achieved an unprecedented level of precision for the calibration of jets. The excellent capabilities of the LHC detectors, as well as their accurate simulation software, enabled the development of sophisticated input signals for the reconstruction and calibration of jets using finely segmented calorimeter and high precision tracking information. Inputs to jet reconstruction (topological clustering in ATLAS and particle flow reconstruction in CMS) have a key role in the measurement of jets at the LHC, in particular to mitigate the effects of pileup and to improve the accuracy of the jet energy measurements.  The high integrated luminosity provided by the LHC has also allowed the use of large calibration datasets for {\it in situ} jet energy measurements that significantly reduced the uncertainty on the jet energy scale determination. Lastly, two key ideas from the theory community have significantly influenced the measurement and calibration of jets at the LHC: the early adoption of the anti-$k_t$ jet algorithm before the start of Run 1, and the use of event-by-event pileup subtraction methods to reduce the effect of pileup fluctuations and improve the jet energy resolution. 

This review provides an overview of the methods developed by ATLAS and CMS to reconstruct and calibrate jets at the LHC. Particular emphasis is given to the definition of the input signals used to reconstruct jets. Inputs to jet reconstruction differ greatly between ATLAS and CMS, and these differences can be understood by their different detector characteristics. Section \ref{sec:detector} describes and compares the main features of the ATLAS and CMS detectors relevant for jet reconstruction and calibration. The Monte Carlo samples and datasets used are described in Section \ref{sec:datamc}. Section \ref{sec:overview} provides an overview of the jet energy calibration procedure highlighting the main differences between ATLAS and CMS. Section \ref{sec:reco} focuses on the input constituents used for jet reconstruction and how detector designs and technologies influenced the different choices made by ATLAS and CMS. Lastly, the jet energy calibration methods and the final jet energy scale uncertainty are described in Section \ref{sec:JES}. 

%-----------------------------------------------------------
\section{The ATLAS and CMS detectors }	
\label{sec:detector}
%-----------------------------------------------------------

\begin{figure}[h]
\centering
  \subfigure[]{\includegraphics[width=8.1cm, angle=-90]{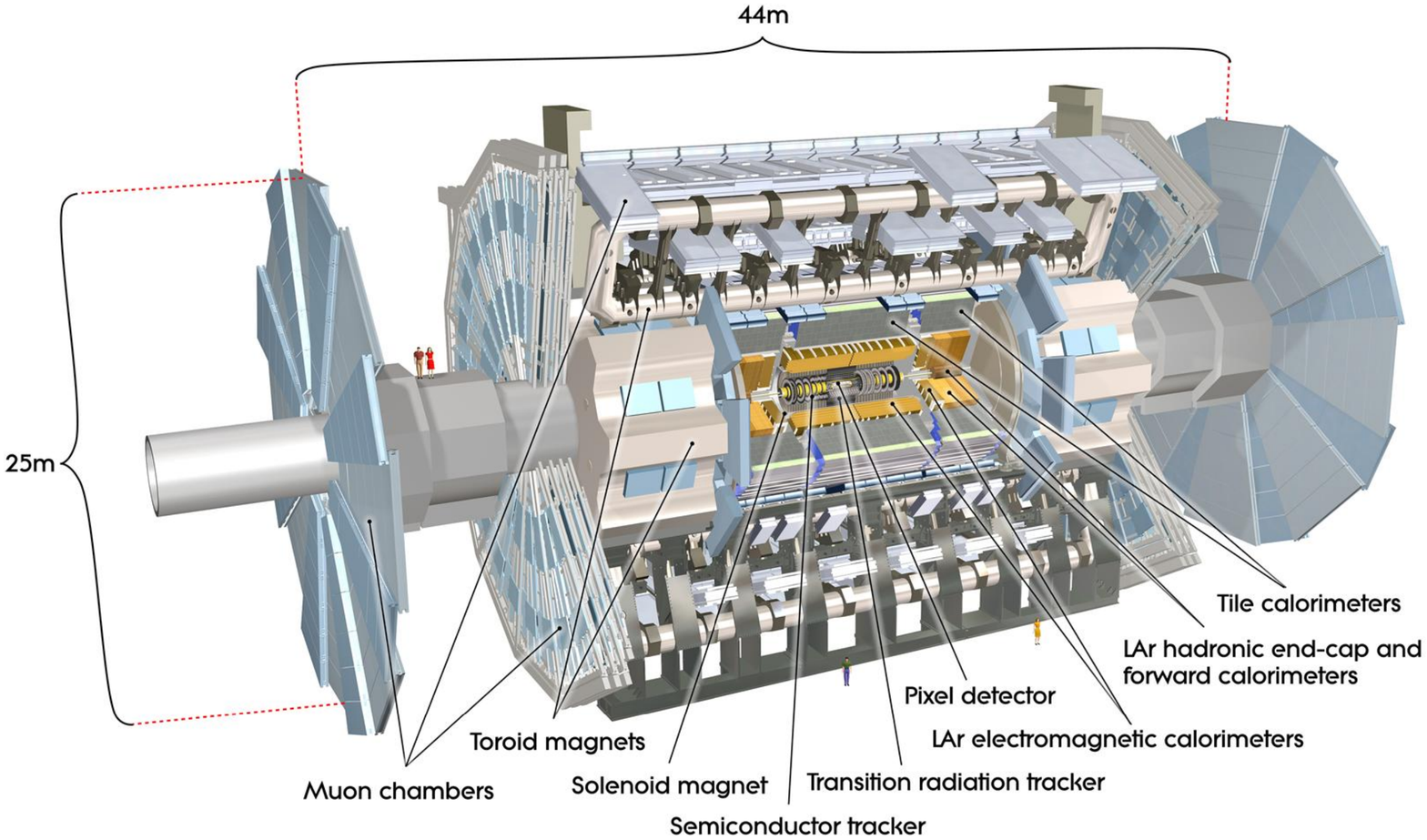} \label{fig:ATLAS}}
  \subfigure[]{\includegraphics[width=8.1cm, angle=-90]{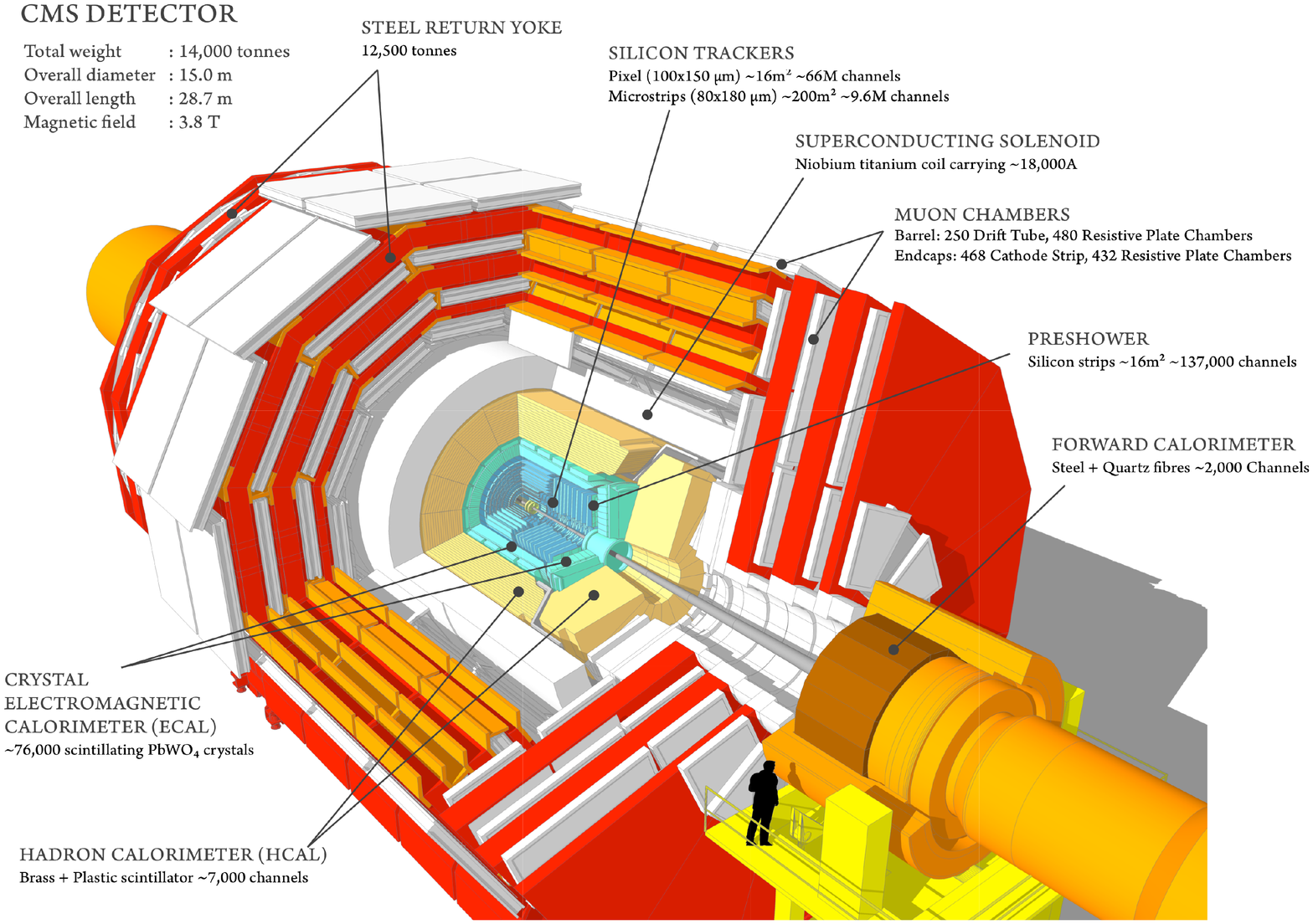} \label{fig:CMS}}
  \caption{Schematic views of the ATLAS  \subref{fig:ATLAS}  and CMS \subref{fig:CMS} detectors. Figures taken from Ref. \protect\refcite{figATLAS} and \protect\refcite{figCMS}.}  
\label{fig:DETECTORS}
\end{figure}

A detailed description of the ATLAS and CMS detectors can be found in Refs. \refcite{ATLASdet} and \refcite{CMSdet}. The ATLAS detector is illustrated in Fig. \ref{fig:ATLAS}. It consists of a tracking system (Inner Detector) with a coverage of $|\eta|<2.5$\footnote{Both ATLAS and CMS use a right-handed coordinate system with its origin at the nominal interaction point (IP) in the center of the detector. The positive x-axis points to the center of the LHC ring, the positive y-axis points upward, and the z-axis is defined parallel to the anticlockwise beam direction. The azimuthal angle $\phi$ is measured with respect to the x-axis in the xy-plane and the polar angle $\theta$ is measured with respect to the z-axis. Pseudo rapidity is defined as $\eta = - \ln [\tan(\frac{\theta}{2})]$. Rapidity is  defined as $y = 0.5\ \ln[(E + p_z)/(E - p_z)]$, where $E$ is the energy and $p_z$ is the $z$-component of the momentum. Transverse energy is defined as $E_T = E \sin \theta$.}, electromagnetic and hadronic sampling calorimeters covering $|\eta|<4.9$, and a muon spectrometer with coverage $|\eta|<2.7$. The Inner Detector is comprised of a silicon pixel detector, a silicon microstrip detector, and a transition radiation tracker detector, all immersed within a solenoid magnet that provides an axial magnetic field of 2T. Charged particles tracks are reconstructed with $p_T>400$ MeV. The calorimeter system is composed of several subdetectors. A highly-segmented liquid-argon (LAr) sampling electromagnetic calorimeter with lead absorber plates is split into a barrel and two end-caps covering the regions $|\eta|<1.475$ and $1.375<|\eta|<3.2$ respectively. The hadronic calorimeter consists of a barrel and an extended barrel sampling calorimeter using steel and scintillating tiles in the range $|\eta|<1.7$ (Tile). The hadronic end-cap calorimeter uses copper/LAr technology covering the range $1.5<|\eta|<3.2$. A copper-tungsten/LAr forward calorimeter (FCAL) extends the coverage up to $|\eta|=4.9$. The LAr barrel, end-cap, and forward calorimeters have a long charge collection time (typically 400-600 ns) and are sensitive to up to 24 (12) bunch crossings during nominal (Run 1) 25 ns (50 ns) spacing. The Tile calorimeter has a faster response and is less sensitive to out-of-time signals. The LAr calorimeters have three longitudinal layers in the barrel, four in the end-cap, and three in the FCAL. The Tile calorimeter has three longitudinal layers and additional scintillator detectors in the gap region between the Tile barrel and the extended barrel. In addition, there is a pre-sampler layer in front of the LAr calorimeter. The transverse segmentation of the LAr calorimeter varies between $(\eta \times \phi) = (0.025 \times 0.025)$ and $(\eta \times \phi) = (0.1 \times 0.1)$ depending on the longitudinal layers and $\eta$. The Tile and FCAL calorimeters have a transverse segmentation between $(\eta \times \phi) = (0.1 \times 0.1)$ and $(\eta \times \phi) = (0.2 \times 0.2)$. The combined depth of the calorimeters for hadronic energy measurements is larger than 10 hadronic interaction lengths ($\lambda$) across the full detector acceptance. Surrounding the ATLAS calorimeters, the muon system is comprised of three air-core toroids, a barrel and two end-caps, generating a magnetic field in the pseudorapidity range of $\eta<2.7$, and a muon spectrometer with three layers of precision tracking chambers. The trigger system consists of three levels. A hardware-based level 1 (L1) reduces the event rate to 75 kHz and it is followed by two software-based high-level triggers (HTL) which reduced the event rate to about 400 Hz. 

The CMS detector is shown in Fig. \ref{fig:CMS}. It consists of a tracking system comprised of a silicon pixel and a silicon microstrip detector covering the region $|\eta|<2.5$, a high-granularity PbWO$_4$ crystal electromagnetic calorimeter (ECAL), and a brass/scintillator hadronic calorimeter (HCAL) with a coverage of $|\eta|<3$, all inside a 3.8T axial magnetic field provided by a superconducting solenoid magnet.  A forward iron/quartz-fiber hadronic calorimeter covers the region $3<|\eta|<5$, outside the magnetic field volume.  A lead and silicon-strip preshower end-cap detector is placed in front of the ECAL calorimeter. The ECAL calorimeter has a transverse granularity of $(\eta \times \phi) = (0.0174 \times 0.0174)$, whereas the HCAL is 5 times coarser. The muon detector system is outside the solenoid field and uses the steel return yoke and gaseous detectors to identify muons up to $|\eta|<2.4$. The CMS HCAL calorimeter has a fast charge collection time such that it is primarily sensitive to signals produced within 2 bunch crossings. The pulse shape of the ECAL calorimeter is longer and requires approximately 10 bunch crossings to collect 95\% of the energy of incident particles. Charged particle tracks can efficiently be reconstructed with a small fake rate down to a transverse momentum ($p_T$) of 150 MeV.  The trigger system is comprised of a first level hardware-based trigger (L1) that reduces the event rate to about 100 kHZ, and a high-level software-based trigger (HLT) that reduces the event rate to less than 1 kHz. 

There are several distinctive characteristics between the ATLAS and CMS detectors that are critical to understand the different approaches to jet reconstruction adopted by each collaboration, as well as to understand the differences in jet performance. The ECAL and HCAL calorimeters in CMS are immersed inside a very high magnetic field, whereas the ATLAS calorimeter system is outside a magnetic field that has an intensity of almost a factor of two smaller. The high magnetic field, in combination with the very fine transverse granularity of the ECAL, allows for a good separation between photons and hadronic showers, a key feature that will be exploited by the CMS particle flow event reconstruction.  ATLAS, on the other hand, has significant material in front and between the calorimeter systems due to the solenoid itself as well as the LAr cryostat before the Tile calorimeter. The high magnetic field in CMS also enables the reconstruction of very low $p_T$ tracks. However, since more charged particles will be bent out by the magnetic field not reaching the calorimeter, CMS will need to rely on a more integrated use of tracks and calorimeter signals for jet reconstruction to recover these low $p_T$ signals. ATLAS calorimeters have a fine longitudinal segmentation, which, combined with their fine transverse segmentation and excellent hadronic energy resolution, will enable the use of 3-dimensional clustering algorithms exploiting depth and shape information to identify and calibrate clusters initiated by hadronic and electromagnetic showers, and correct their energies for dead material effects.  The ATLAS detector, hence, has more handles for calorimeter-based jet reconstruction and calibration whereas CMS has more capabilities for the integration of calorimeter and tracking information at the level of input signals to jet reconstruction. Both detectors are non-compensating, meaning that the calorimeter response to hadronic and electromagnetic particles is different. Non-compensation results in increased energy fluctuations, leading to a degraded jet energy resolution, and a non-linear and a flavor-dependent jet energy response. Compensating for these effects will require specific methods to reconstruct and calibrate the input signals to jet reconstruction. Section \ref{sec:reco} will discuss the different solutions implemented by ATLAS and CMS to correct for these effects, exploiting the different strengths and main features of their detectors. 

%-----------------------------------------------------------
\section{Data and Monte Carlo samples}
\label{sec:datamc}
%-----------------------------------------------------------

This review considers data selected from the full 2012 p-p data taking period at a center-of-mass energy of 8 TeV, for an integrated luminosity of approximately 20fb$^{-1}$. Monte Carlo event generators are used to simulate jets produced in p-p collisions. ATLAS uses \PythiaP \cite{ref:pythia} with the ATLAS A2 tune \cite{ref:A2} and the CT10 \cite{ref:CT10} NLO parton distribution function (PDF) set as the primary MC event generator to derive the jet energy calibration. Samples used to determine systematic uncertainties are based on the \HerwigPP MC event generator \cite{ref:herwig1, ref:herwig2}. Additional samples used for the determination of the jet energy scale are described in Ref. \cite{ref:datadriven}. Pileup is simulated by overlaying additional p-p collisions generated with \PythiaP. These events are overlaid onto the hard scattering events following a Poisson distribution around the measured average number of additional p-p collisions per bunch crossing, $\mu$. The simulation also accounts for pileup events occurring in nearby bunch crossings. Generated events are passed through a full simulation of the ATLAS detector \cite{ref:ATLAS19} based on \geant \cite{ref:G4} that simulates the interactions of the particles produced by the event generators after parton shower and hadronization with the detector material. Hadronic showers are simulated with the QGSP BERT model \cite{ref:QGSP}. CMS uses \PythiaS \cite{ref:pythia6} with the tune Z2* \cite{refZ2} as its main MC event generator. Additional samples used for systematic uncertainties are generated with \HerwigPP, tune EE3C. For the determination of the jet energy scale in \Zb jet events, the \Madgraph \cite{ref:madgraph} generator interfaced with \PythiaS is used to simulate top pair, W+jets and Drell-Yan production and the \Powheg \cite{ref:powheg} generator interfaced with \PythiaS for single top production. Monte Carlo samples are propagated through the CMS detector simulation based on \geant. 

%-----------------------------------------------------------
\section{Overview of jet energy calibration at the LHC}
\label{sec:overview}
%-----------------------------------------------------------

The purpose of the jet energy calibration is twofold. First, the energy scale of reconstructed jets does not correspond to the {\it truth-particle jet} energy scale (JES), defined as the energy of jets built from all stable Monte Carlo particles from the hard interaction only, including the underlying event (UE) activity. A dedicated jet energy calibration is then needed to calibrate, on average, the reconstructed jet energy to that of the corresponding truth-particle jet. The energy scale calibration needs to also correct for the effect of pileup. Second, the jet energy calibration has to bring the energy scale of jets in data and simulation to the same footing.  

A key figure of merit to measure and characterize the jet energy scale is the jet energy {\it response}, defined as the ratio between the reconstructed jet energy and the corresponding truth-particle jet energy in the simulation: 

\begin{equation}
R(E, \eta) = \Big \langle \frac{E^{reco}}{E^{truth}} \Big \rangle
\label{RESPONSE}
\end{equation}

Sometimes transverse momentum $p_T$ is used to define the jet response. There are several effects that cause the jet energy response to be smaller than 1 that need to be accounted for in the JES calibration. Calorimeter non-compensation due to the difference in response between electromagnetic and hadronic showers, energy losses in inactive regions of the detector, energy deposits below noise thresholds or energy left outside the input signals used to build jets, energy of truth particles not included in the jet, and longitudinal energy leakage from showers not fully contained by the calorimeters. In addition, pileup introduces major effects affecting the jet energy response as well as its resolution. Pileup adds an energy {\it offset} to jets due to the contribution of additional pileup particles inside jets. These additional particles increase the jet energy response and also make it luminosity-dependent. %Furthermore, fluctuations of pileup within events and from one event to another introduce fluctuations that degrade the jet energy resolution and create fake pileup jets from particles not belonging to the main hard-scatter (HS) interaction. 

Solutions to these effects are accomplished by a combination of complex input signal algorithms and calibrations before jet reconstruction, and by a sequence of JES corrections applied to reconstructed jets and derived from data and simulation. Both ATLAS and CMS have developed similar strategies for the JES calibration. However, they significantly differ in their choice of input constituents to jet reconstruction and their calibration. The definition and calibration of input signals before jet reconstruction plays a major role at the LHC and will be discussed extensibly in section \ref{sec:reco}. Input signals in ATLAS ({\it topoclusters}) and CMS ({\it particle flow}) were designed to partially correct for pileup, calorimeter non-compensation, and dead material effects, leaving jet-level effects such as out-of-cone, thresholds, and data to Monte Carlo differences to the JES correction.

%-----------------------------------------------------------
\section{Jet reconstruction}
\label{sec:reco}
%-----------------------------------------------------------

Jets are reconstructed combining signals from different sub-detectors, primarily from the calorimeters and tracking subsystems. ATLAS exploits its fine longitudinal and transverse segmentation as well as its excellent hadronic calorimeter resolution to build topological clusters from calorimeter cells (topoclusters). Calorimeter-only topological clusters are the primary input to jet reconstruction. After jet finding, ATLAS utilizes tracking information to refine the calibration of jets, and to further suppress pileup effects. CMS uses a different concept where tracks and calorimeter signals are used in combination to build the basic input signals to jet reconstruction in the so called particle flow paradigm. This approach takes advantage of key CMS detector features such as its high magnetic field and excellent tracking and electromagnetic calorimetry. Despite these very different approaches, mainly driven by detector considerations, both experiments integrate tracking with calorimeter information for optimal performance. 

\subsection{ATLAS topological clustering}
\label{sec:topo}

Topoclustering \cite{ref:topo} is a three-dimensional clustering algorithm at the level of individual calorimeter cells. It incorporates two key features: a built-in noise suppression mechanism to limit the formation and grow of clusters from electronic and pileup noise, and a {\it local cluster weighting} scheme (LCW) to classify clusters as hadronic-like or electromagnetic-like and calibrate them appropriately.  The former provides, effectively, a constituent-level pileup suppression that reduces the pileup contributions before jet reconstruction, while the latter significantly improves the jet energy resolution and reduces the flavor dependence of the response. Topological clusters are formed by connecting calorimeter cells using a three-dimensional nearest-neighbor algorithm. Clusters are built grouping cells with energy significance $|E_{\rm cell}|/\sigma^{\rm noise}>4$ for the seed, $|E_{\rm cell}|/\sigma^{\rm noise}>2$ for the neighbors, and $|E_{\rm cell}|/\sigma^{\rm noise}>0$ at the boundary. $\sigma^{\rm noise}$ is the total cell noise which is the sum in quadrature of the cell electronic noise and the noise from pileup ($\sigma_{pile-up}^{\rm noise}$).  The cell pileup noise contribution is determined from the simulation and depends on the average number of interactions per crossing (\avgmu) and the bunch spacing $\Delta t$.  The value of $\sigma^{\rm noise}$ effectively adjusts the energy significance thresholds that determine the seeding and growth of clusters. Since cluster formation and calibration, as well as jet energy calibration, strongly depend on the value of $\sigma^{\rm noise}$, this parameter is fixed for three specific data taking periods (see Fig. \ref{fig:pileup}): $\sigma^{\rm noise}(\mu=0)$ in 2010, $\sigma^{\rm noise}(\mu=8)$ in 2011, and $\sigma^{\rm noise}(\mu=30)$ in 2012. 
%Figure~\ref{fig:pileupnoise} shows the electronic and total noise per cell as function of $\eta$ and longitudinal depth for two distinct data taken periods in which $\sigma^{\rm noise}$ was adjusted. 
%
%\begin{figure}[b]
%\centering
% \subfigure[$\sigma^{\rm noise}(\mu=0)$ in 2010]{
%  \includegraphics[width=6.0cm]{noise_tot_plot_OFLCOND-SDR-BS7T-04-09} \label{fig:2010}}
% \subfigure[$\sigma^{\rm noise}(\mu=30)$ in 2012]{
%  \includegraphics[width=6.0cm]{noise_tot_plot_OFLCOND-MC12-SDR-06} \label{fig:2012}}
%\caption{Cell noise as a function of $\eta$ and calorimeter longitudinal layer for \subref{fig:2010} the 2010 configuration with $\avgmu=0$, and \subref{fig:2012} the 2012 configuration with $\avgmu=30$. The colors correspond to the noise in the pre-sampler (PS) and the up to three layers of the LAr EM calorimeter, the up to three layers of the Tile calorimeter, the four layers for the hadronic end-cap (HEC) calorimeter, and the three modules of the forward (FCAL) calorimeter.}
%\label{fig:pileupnoise}
%\end{figure}
After topoclusters are found, a splitting algorithm further separates clusters based on local energy maxima within clusters. The topoclusters are first reconstructed at the {\it EM} scale \cite{ref62,ref66,ref67} which is defined as the calibrated energy scale for electromagnetic particles. In a second step, the probability that a topocluster is generated by an electromagnetic shower is computed using local cell and cluster level information. This step is called cluster classification and is performed based on the cell energy density and the longitudinal shower depth of the clusters. Topoclusters then receive electromagnetic and hadronic calibrations based on their classification probability. Calibrations are derived from single pion Monte Carlo samples in events with no pileup and correct for calorimeter non-compensation, energy losses in non-instrumented regions and losses due to noise threshold effects, and cluster non-isolation \cite{ref3}. Calibrated topoclusters are referred to as {\it LCW} ({\it Local Cluster Weighted}) topoclusters.

\subsection{CMS particle flow}

The particle flow event reconstruction attempts to reconstruct all stable particles individually, combining information from multiple sub-detectors. Measurements of jet fragmentation at LEP \cite{LEP1} as well as Monte Carlo simulations at the LHC showed that, on average, about 60\% of the jet energy is carried by charged particles, 30\% by photons, and 10\% by neutral hadrons. In the particle flow formalism, the momentum of charged particles is measured with very high precision in the tracking detector, photons are measured with the high granularity and high resolution electromagnetic calorimeter, and leptons with a combination of tracking and calorimeter information. Hence, only the relatively small fraction of the jet energy carried by neutral hadrons relies on the hadronic calorimeter with coarser segmentation and poorer resolution. The key to particle flow event reconstruction and what determines its performance is the ability to separate individual showers from charged particles and photons, which is primarily limited by the segmentation and inner radius of the electromagnetic calorimeter, the magnitude of the solenoid magnetic field, and the amount of material in front of the calorimeter system. The success of the particle flow algorithm in CMS greatly depends on these specific detector features. While particle flow algorithms have been used at previous collider experiments \cite{refLEP}, the CMS particle flow algorithm is the first technique to be optimized in a high pileup environment.   Not only the performance of particle flow energy reconstruction is maintained at high luminosity, by exploiting the vertex position information of tracks, the CMS particle flow algorithm has been effective at reducing the effect of pileup by removing objects originating from pileup vertices. This new aspect of particle flow is called {\it Charged Hadron Subtraction} (CHS) and it is described in section \ref{sec:pileup}.

\subsection{Jet algorithms}

The main jet algorithm used by both ATLAS and CMS experiments at the LHC is the {\it anti-$k_t$}\cite{refAKT} sequential recombination algorithm with the four-momentum recombination scheme. Both collaborations use the FastJet software \cite{refFASTJET} to reconstruct jets.  The analysis of boosted topologies using jet substructure was performed using a wide range of jet and grooming algorithms but it is outside the scope of this review.  The anti-$k_t$ algorithm provides several experimental advantages that led to its early adoption by both detector collaborations. Most notably, its reduced sensitivity to pileup compared to the $k_t$ and $C/A$ algorithms \cite{refAKT2}, and its regular, circular, shape that facilitates its energy calibration.  The choice of the R parameter was different between ATLAS and CMS. ATLAS (CMS) utilized R=0.4 (0.5) and R=0.6 (0.7). 

The inputs to jet reconstruction are positive energy topoclusters (particle flow objects) in the case of ATLAS (CMS). Truth-particle jets are built from truth Monte Carlo particles from the hard-scatter interaction only, including the underlying event. In this case, the inputs to jet reconstruction are stable particles\footnote{ATLAS defines a particle as stable if it has a lifetime of at least 10 ps while CMS requires $c \tau >$ 1cm} excluding muons and neutrinos in the case of ATLAS, and excluding only neutrinos in CMS. Truth-particle jets are constructed to provide a reference for the calibration of reconstructed jets. Truth-particle jets in ATLAS do not include muons because jets are built from calorimeter-only topoclusters, excluding muons. 
%Truth-particle jets represent the measurement of an ideal detector, without pileup, but including the underlying event. 

%-----------------------------------------------------------
\section{Jet energy calibration}
\label{sec:JES}
%-----------------------------------------------------------

\begin{figure}[h]
\centering
  \subfigure[]{\includegraphics[width=12.0cm]{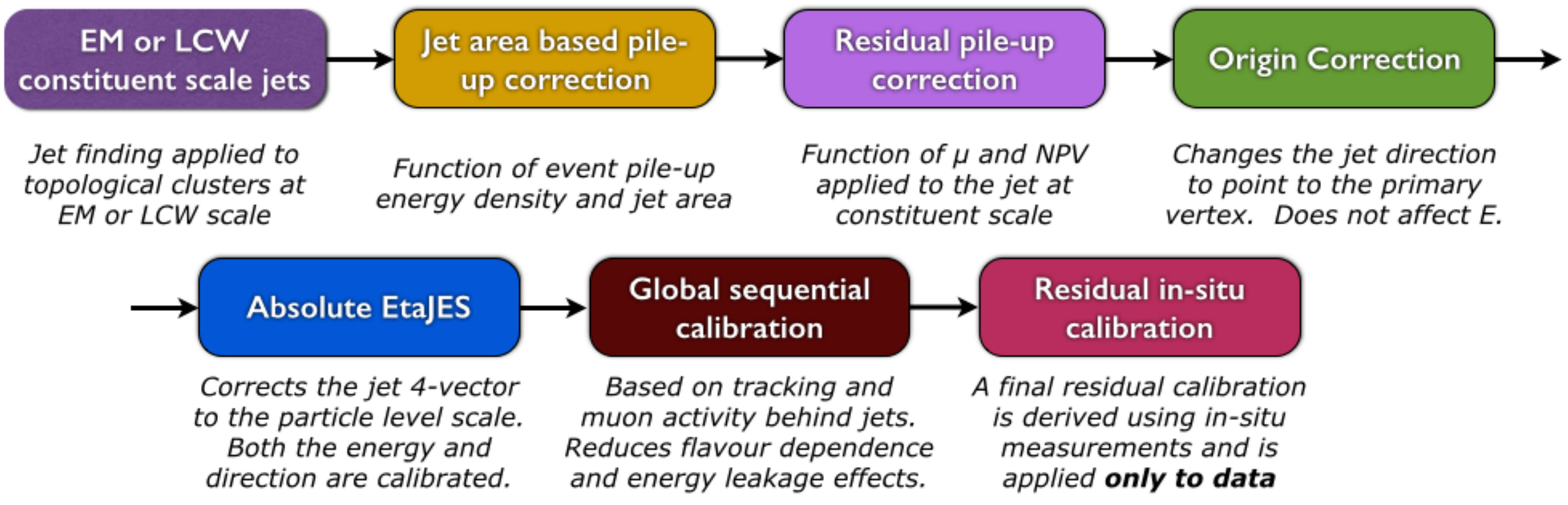} \label{JESATLAS}}
  \subfigure[]{\includegraphics[width=12.0cm]{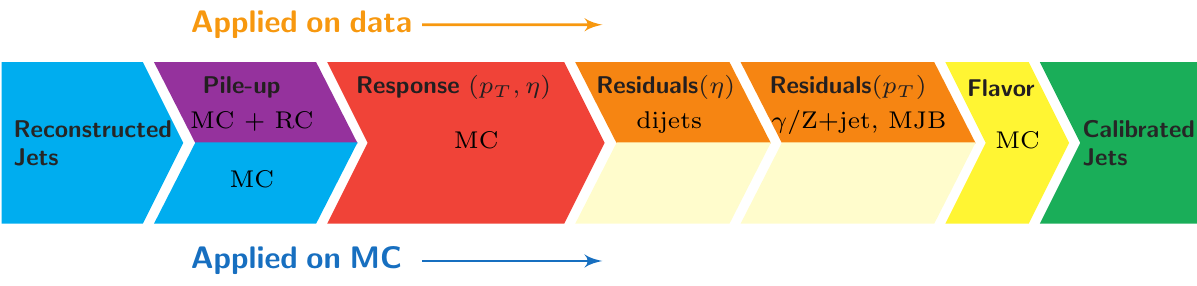} \label{JESCMS}}
  \caption{Overview of the ATLAS  \subref{JESATLAS} and CMS \subref{JESCMS} jet calibration schemes. Figures taken from Ref. \protect\refcite{ref:ATLAScombo} and \protect\refcite{ref:CMScombo}.}  
\label{fig:JESSCHEMES}
\end{figure}

Jets are calibrated to truth-particle level using a factorized, sequential scheme, consisting
of several steps. At very high level, both ATLAS and CMS use a similar strategy consisting of three main components. The first step is an additive offset pileup correction to remove the effect of additional energy from pileup particles inside the jet. Both collaborations utilize event-by-event pileup corrections to reduce pileup fluctuations from one event to the next. This correction makes the jet response independent of the number of primary vertices in the event. The next step is the application of a multiplicative jet energy scale correction derived from MC events. The goal of this correction is to restore the jet response to that of truth-particle jets in QCD dijet events. The last step is a residual {\it in situ} correction that is only applied to jets in data.  This residual correction, computed as the ratio of MC to data jet energy response, brings the energy response of jets in data and MC to agreement, reducing the jet energy scale systematic uncertainty. In the absence of a residual in situ correction, a major source of jet energy scale uncertainty would come from the difference in energy response between jets in data and simulation. With the residual correction, instead, the jet energy scale uncertainty is now given by the uncertainty on the measurement of the jet response in data.  At the LHC, large samples of dijet, \Zjets, and \gj~ events have enabled in situ measurements of the jet response with very high precision, significantly reducing the JES uncertainty. Optional, post-calibration jet-by-jet corrections are available, in particular  using tracking information in the case of ATLAS.  Figure \ref{fig:JESSCHEMES} shows an overview of the jet energy calibration procedure in ATLAS and CMS.  

The general features of the JES calibration procedure in ATLAS and CMS are the same. ATLAS has two additional corrections that are not present in CMS: an origin correction that corrects the calorimeter jet direction pointing back to the primary vertex position rather than to the nominal center of the ATLAS detector, and a {\it Global Sequential Calibration} (GSC) that applies additional JES corrections using tracking and jet shape information reducing the flavor dependence of the response and improving the jet energy resolution. These two corrections are not present in CMS because these aspects are accounted for in the context of particle flow reconstruction. CMS also has a dedicated and optional flavor response correction.

%-----------------------------------------------------------
\subsection{Pileup}
\label{sec:pileup}

The high instantaneous luminosity achieved by the LHC resulted in unprecedented rates of multiple simultaneous p-p interactions within single bunch crossing (pileup). Pileup poses one of the most difficult challenges for the reconstruction and calibration of jets at the LHC.  Multiple p-p interactions within the same event (in-time pileup) create additional particles that contaminate jets. The in-time pileup jet energy contamination is proportional to the area of the jet. Pileup interactions can also create {\it fake} pileup jets. Pileup jets can originate from any of the additional hard p-p interactions in the event ({\it QCD} pileup jets), or from random combination of soft particles originating from multiple vertices ({\it stochastic} pileup jets).  Calorimeter detectors are also sensitive to collisions occurring in the previous and subsequent bunch crossings (out-of-time pileup). This is due to the relatively long charge collection time in the calorimeters (400-600~ns in the ATLAS LAr calorimeter, and approximately 100~ns in the CMS ECAL) compared to the LHC bunch crossing interval (25~ns design, and 50~ns in Run 1). In ATLAS this sensitivity is reduced using a fast bipolar shaped calorimeter signal with net zero integral over time that leads to a cancellation, on average, of in-time and out-of-time pileup fluctuations \cite{LAR1, LAR2}. CMS ECAL signal reconstruction uses three preceding and five subsequent samples spaced by 25~ns to remove, on average, a varying pedestal.   

Pileup mitigation methods at the LHC are based on three major techniques: constituent-level pileup suppression using topological clustering (ATLAS) and charged hadron subtraction (CMS), event-by-event pileup subtraction using the jet areas method \cite{refareas}, and pileup jet tagging and suppression using tracking and vertex information. Charged tracks provide a crucial tool for reducing the effects of pileup on jets since they can be accurately associated to primary vertices. Tracks originating from pileup vertices can be directly removed before jet reconstruction (CHS in CMS). Charged tracks also provide information to identify and reject pileup jets ({\it Jet Vertex Fraction}~\cite{JVF} in ATLAS and {\it Pileup Jet ID}~\cite{PileupID} in CMS). The jet areas method estimates the pileup activity event-by-event and subtracts it jet-by-jet using the novel concept of {\it jet area}. The event-by-event nature of this correction reduces some of the effects of pileup fluctuations on the jet energy measurement, improving the jet energy resolution. 

One of the main figures of merit to evaluate the effect of pileup in jets and to study the performance of pileup corrections is the jet $p_T$ {\it offset}, defined as the difference between the reconstructed and the truth-particle jet $p_T$ in the simulation: 

\begin{equation}
O = p^{jet}_T - p^{truth}_T
\end{equation}

\subsubsection{Charged hadron subtraction}

\begin{figure}[b]
\centering
  \subfigure[]{\includegraphics[width=5.0cm]{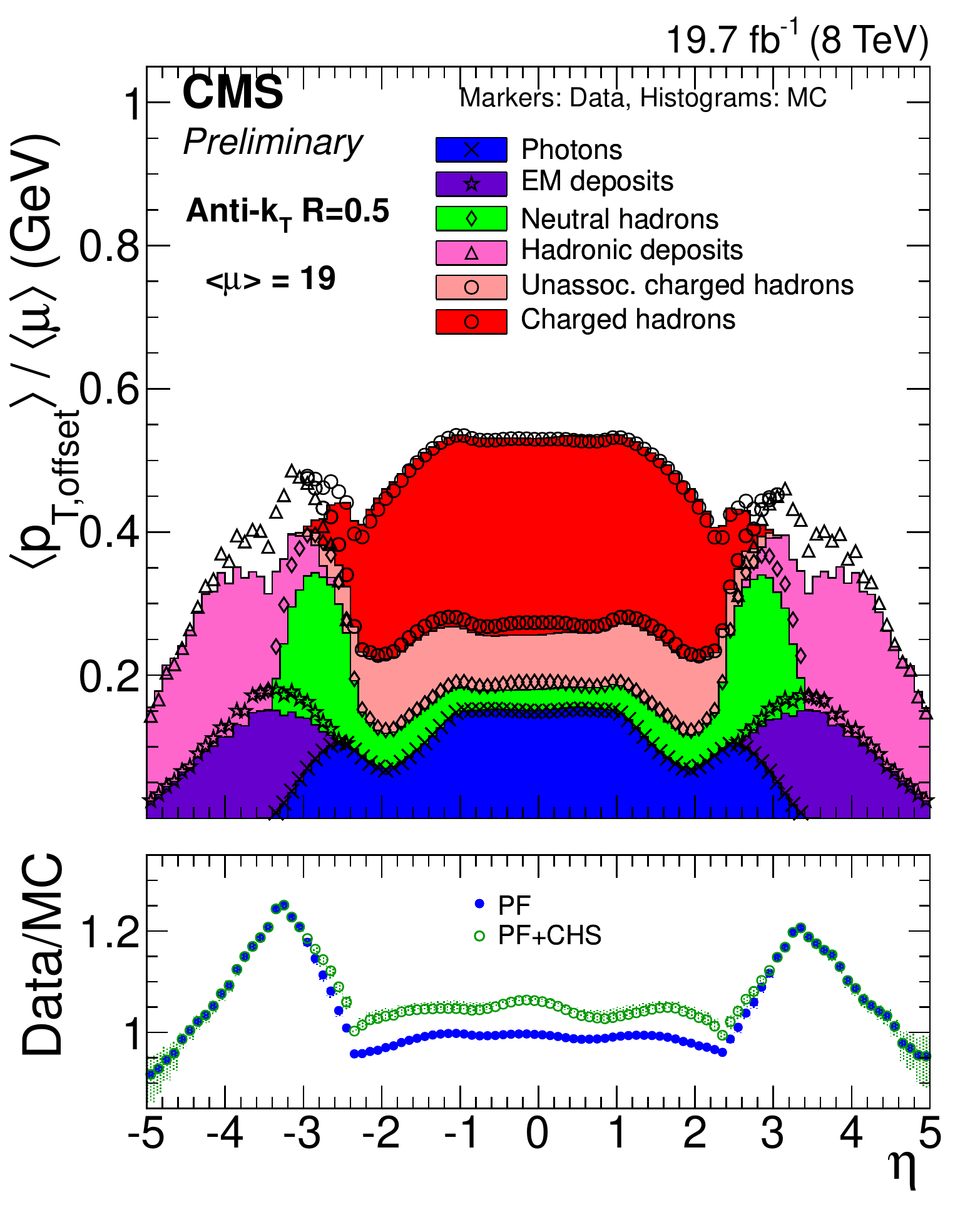} \label{figCHS}}
 \subfigure[]{\includegraphics[width=6.5cm]{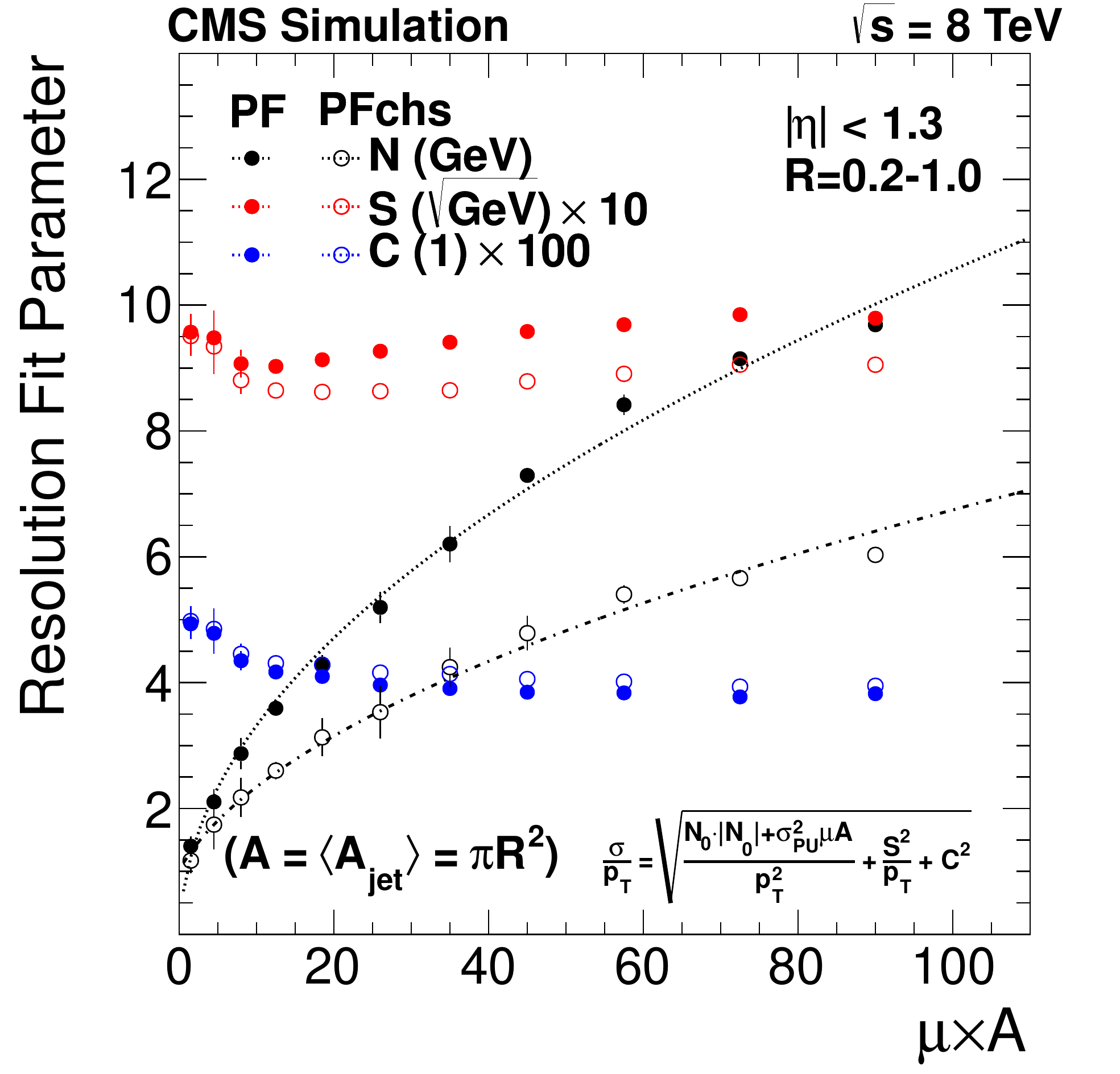} \label{figNOISE}}
 \caption{\subref{figCHS} Average $p_T$ offset as a function of jet $\eta$ for the different particle flow candidates in data and MC. \subref{figNOISE} Parameters of the fits to the fractional jet $p_T$ resolution in the barrel region ($|\eta|<1.3$) as a function of $\mu *A$. The fit parameters are averaged over cone radii $R=0.2$ through $R=1.0$. Figures taken from Ref. \protect\refcite{ref:CMScombo} and \protect\refcite{ref:CMScones}.}
\label{pflowres}
\end{figure}

Charged hadron subtraction reduces the in-time pileup contribution from charged particles by rejecting particle flow candidates originating from well reconstructed pileup vertices before jet reconstruction. After track and vertex quality requirements, CHS removes approximately 70\% of the charged pileup transverse momenta contributing to jets, as shown in Fig.~\ref{figCHS}. The removal of charged track pileup tracks before jet reconstruction significantly reduces the effect of charged particle pileup fluctuations, improving the jet energy resolution at low $p_T$. Figure~\ref{figNOISE} shows the parameters of fits to the fractional jet energy resolution as a function of the average luminosity $\avg{\mu}$ times the jet area for particle flow jets with (open circles) and without (solid circles) CHS. The stochastic ($S$) and constant ($C$) terms are stable with pileup. The noise term ($N$), which has a $\sqrt{\mu \times A}$ dependence from pileup noise contributions, is significantly smaller after the application of the CHS algorithm.   

%%------------------------------    

\subsubsection{Pileup subtraction}

To remove the effect of pileup on the jet energy, an event-by-event subtraction method is employed. The jet areas technique \cite{refareas} uses $\rho$, the median $p_T$ density of each event in the $\eta \times \phi$ plane, and the area of the jet in this plane. The event $p_T$ density $\rho$ provides an estimate of the global (averaged over the whole event) pileup activity in each event, whereas the jet area is a measure of the jet's susceptibility to pileup.  The pileup energy subtraction is computed as an additive correction to the jet $p_T$, as shown in Eq \ref{eq:areas}:
\begin{equation}
 p^{corr}_T = p^{jet}_T - \rho \times A^{jet}
\label{eq:areas}
\end{equation}

%ATLAS and CMS use slightly different parameters for computing $\rho$, optimized based on experimental aspects such as occupancy and noise threshold effects. 
ATLAS computes $\rho$ in the range of $|\eta|<$2.0. This choice is motivated by the change in calorimeter segmentation above $\eta=$2.5 which results in very low topocluster occupancy in the forward region. In the case of CMS, the median $p_T$ density is $\eta$ dependent, based on Monte Carlo predictions. After the application of the jet areas subtraction, a residual small pileup dependence on the jet energy remains. This is due to several experimental effects such as calorimeter occupancy, topocluster noise threshold effects and, in the case of ATLAS, the variations of the pileup density with $\eta$ which is not captured by the global (event based) $\rho$ estimate. A residual pileup correction is applied after Eq.~\ref{eq:areas} to account for all these effects. ATLAS and CMS define this residual correction in a slightly different way, due to different experimental effects and biases that are detector specific and depend on the details of the input constituents used to reconstruct jets. CMS, in particular, introduces a jet $p_T$ dependent residual correction whereas in ATLAS the $p_T$ dependence of the pileup subtraction is accounted for as a systematic uncertainty.    

\begin{figure}[h]
  \centering
    \subfigure[]{\includegraphics[width=6.0cm]{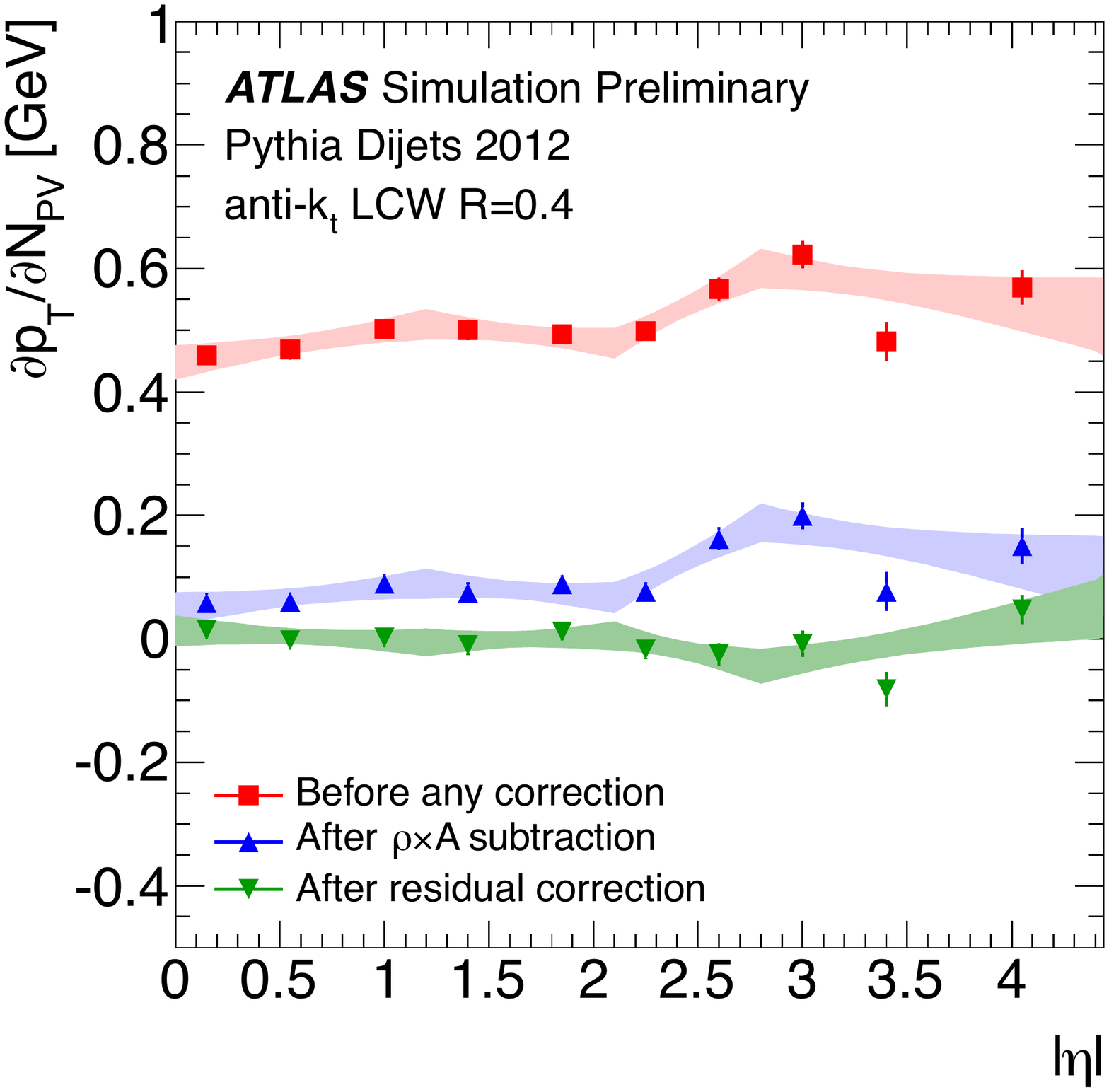}\label{fig:pileup1}}
    \subfigure[]{\includegraphics[width=6.0cm]{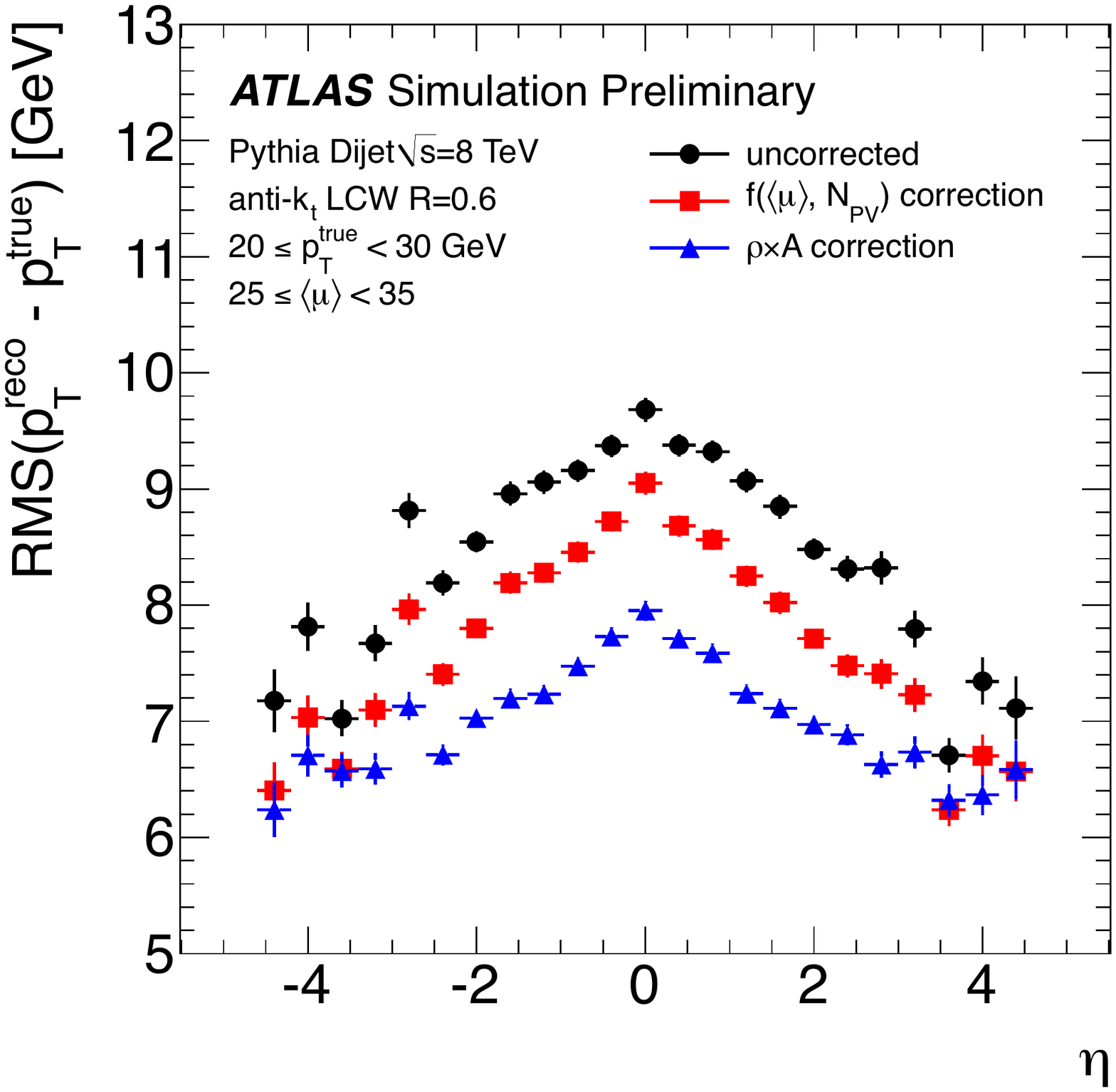} \label{fig:pileup2}}
\caption{\subref{fig:pileup1} Dependence of the reconstructed jet $p_T$ (\antikt\, $R = 0.4$, LCW scale) on in-time pileup  at various correction stages: before any correction, after $\rho\times\jetarea$ subtraction, and after the residual correction.  The error bands show the 68\% confidence intervals of the fits. \subref{fig:pileup2} RMS width of the $\ptreco - \pttrue$ distribution versus pseudorapidity $\eta$ for \antikt\ $R=0.6$ anti-$k_t$ jets matched to truth-particle jets satisfying $20 < \pttrue < 30$~GeV, in simulated dijet events.  A significant improvement is observed compared to a previous subtraction method (shown in red). Figure taken from Ref. \protect\refcite{JVF}.}
\label{fig:subtraction:resolution}
\end{figure}

Figure~\ref{fig:pileup1} shows the dependence of the jet $p_T$ on the number of primary vertices in the event and as a function of $\eta$, before and after pileup subtraction, and after the final residual correction. Before subtraction, each primary vertex adds approximately $0.5$~GeV to each R=0.4 anti-$k_t$ jet. Pileup subtraction reduces this contribution by a factor of 3-5, to less than $200$~MeV per vertex. The residual correction completely removes the pileup dependence. Figure~\ref{fig:pileup2} shows the improvement of event-by-event subtraction on the RMS of the jet offset, which is directly related to the jet energy resolution. The jet areas correction results in approximately a 10\% reduction on the jet-by-jet pileup energy fluctuations. As comparison, a simpler pileup correction based on a parameterization of the pileup contribution to the jet as a function of the number of vertices used in Ref. \refcite{jespaper2011} does not provide the same level of improvement \cite{JVF}.

%%------------------------------    

\subsubsection{Pileup jet suppression}

\begin{figure}[h]
  \centering
    \includegraphics[width=6.0cm]{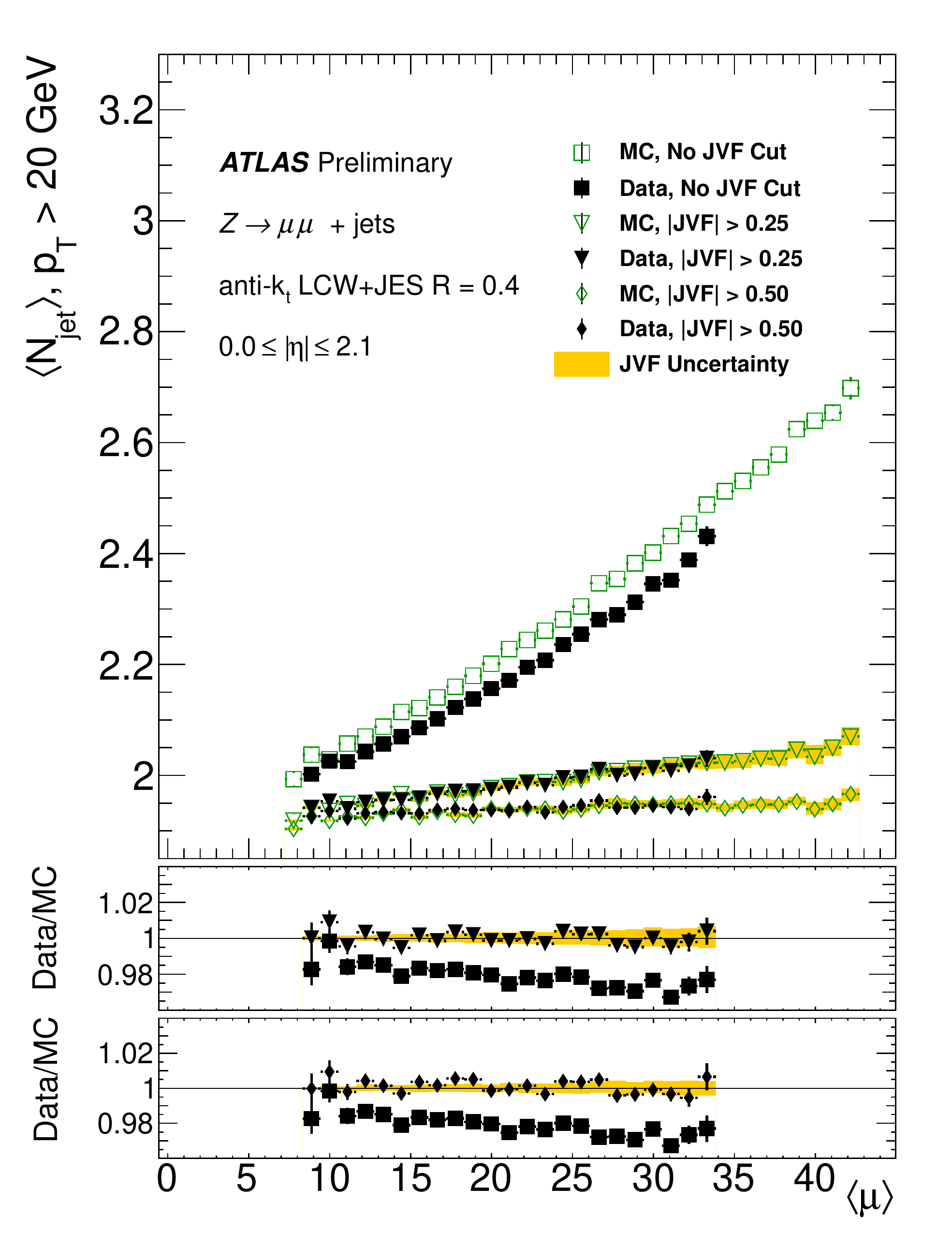}
\caption{The mean anti-$k_t$ R = 0.4 jet multiplicity in data and MC as a function of \avgmu in \Zjets~ events for jets with $\eta <$2.1 before and after a JVF selection cut of 0.25 and 0.5. Figure taken from Ref. \protect\refcite{JVF}.}
\label{fig:JVF}
\end{figure}

Pileup activity can have large local fluctuations within the same event that cannot be captured by the jet areas subtraction. Such fluctuations can give rise to pileup jets that do not originate from the hard scatter interaction and need to be identified and removed. ATLAS and CMS utilize methods to suppress pileup jets that use tracking and vertexing information to identify of the vertex origin of jets. One of such methods is the {\it jet-vertex-fraction} (JVF)~\cite{JVF}, defined as the fraction of track $p_T$ contributing to a jet that originates from the hard scatter vertex. 
%The distribution of JVF is shown in Figure~\cite{fig:JVF1}. Hard scatter jets have large JVF values, whereas jets from pileup tend to have JVF values close to zero. The value -1 is assigned to jets with no associated tracks. 
The application of a JVF selection cut significantly reduces the amount of pileup jets and it also improves the data and MC agreement as it removes a source of fake jets that is generally not well modeled by the simulation. Figure~\ref{fig:JVF} shows the average number of jets above $20$~GeV as a function of the pileup activity before and after the application of two different JVF selections in data and simulation. Before JVF, there is a significant increase in the number of jets due to pileup which is not well described by the MC. After JVF, the average jet multiplicity becomes stable with pileup and MC becomes in agreement with data. More recently, ATLAS introduced an improved pileup suppression technique, the {\it jet-vertex-tagger} (JVT) algorithm, described in Ref. \refcite{ref:JVT}. Pileup jet suppression in CMS is achieved using a boosted decision tree (BDT) discriminant constructed from twelve variables, four of which are based on tracking information. This method is known as {\it pileup jet identification} \cite{PileupID}.

The suppression of pileup jets outside the tracker acceptance is more challenging and less powerful due to the lack of jet-vertex pointing information. CMS utilizes a multivariate discriminant based on jet shape variables such as jet radial moments~\cite{PileupID}. The technique is based on the fact that pileup jets that are stochastic in origin tend to have a broader and more uniform distribution of energy in the transverse plane than QCD jets. ATLAS developed a different approach, which aims at removing QCD rather than stochastic forward pileup jets. The technique uses tracks in the central region to indirectly tag and reject forward pileup jets that are back-to-back to central pileup jets\cite{fJVT}.

\subsubsection{Pileup uncertainties}

Pileup systematic uncertainties are evaluated using in situ validation techniques. In ATLAS two methods\cite{JVF} are used to compare the pileup offset in data and simulation. The first method uses track-jets to provide a reference of the jet $p_T$ that is independent of pileup. The second method exploits the $p_T$ balance between a jet and a $Z$ boson. The total uncertainty, given as a function of jet $p_T$ and $\eta$ is determined by the maximum difference between data and simulation and it is less than 2.5\% for jets with $p_T=20$~GeV. CMS estimates an offset scale factor as the ratio of the jet offset in data and MC using the random cone method in zero bias events~\cite{CMSsubtraction}. The scale factor varies between 5\% for $\eta<2.4$ up to 20\% outside the tracking coverage. The total pileup uncertainty has two components: the uncertainty on the offset scale factor, and the uncertainty on the $p_T$ dependence of the corrected jet areas subtraction. The total uncertainty is approximately 2\% for jet $p_T=20$~GeV, decreasing rapidly with $p_T$.

\subsection{Jet energy scale}

\begin{figure}[h]
  \centering
    \subfigure[]{\includegraphics[width=7.0cm]{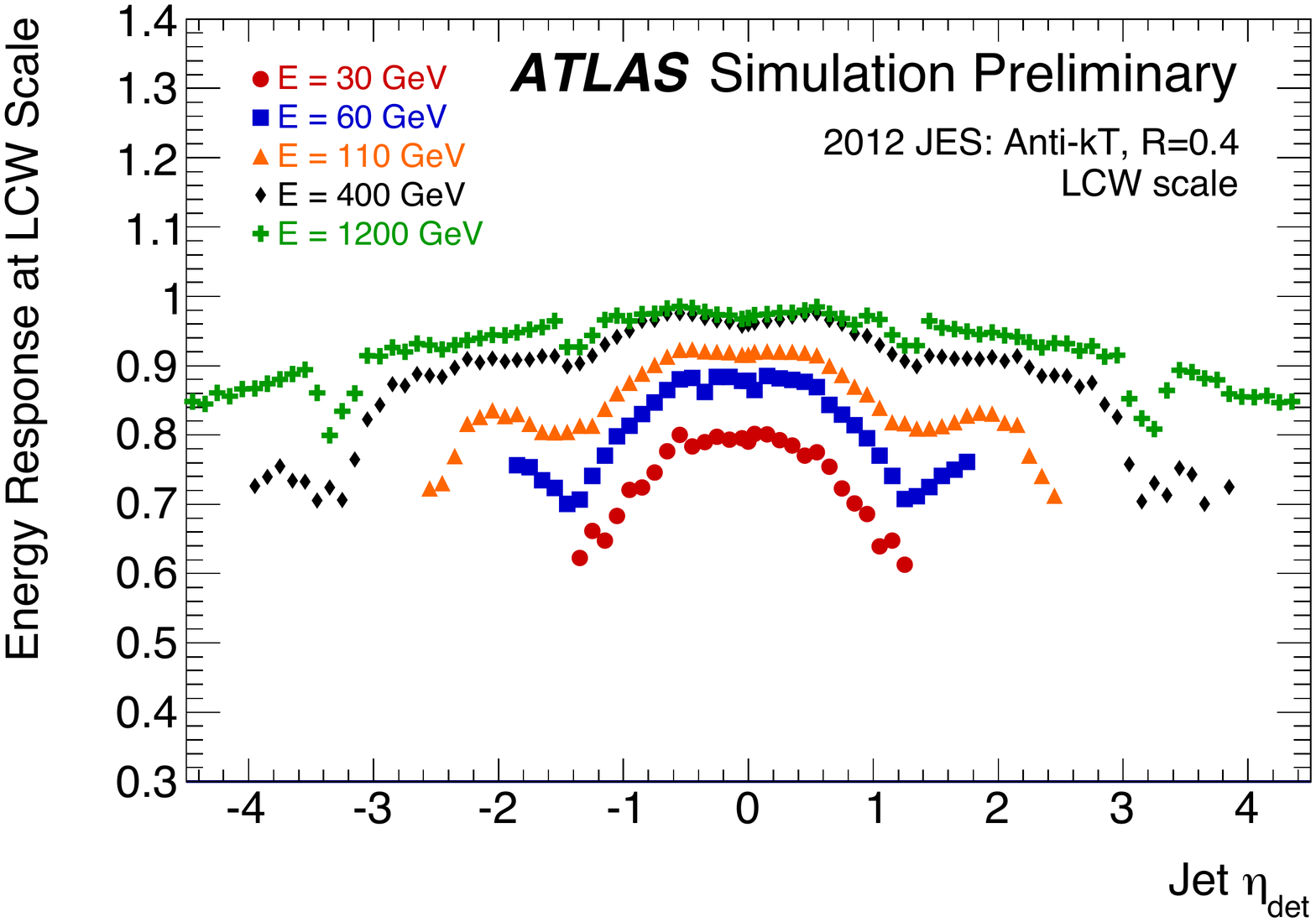} \label{JES_ATLAS}}
    \subfigure[]{\includegraphics[width=5.0cm]{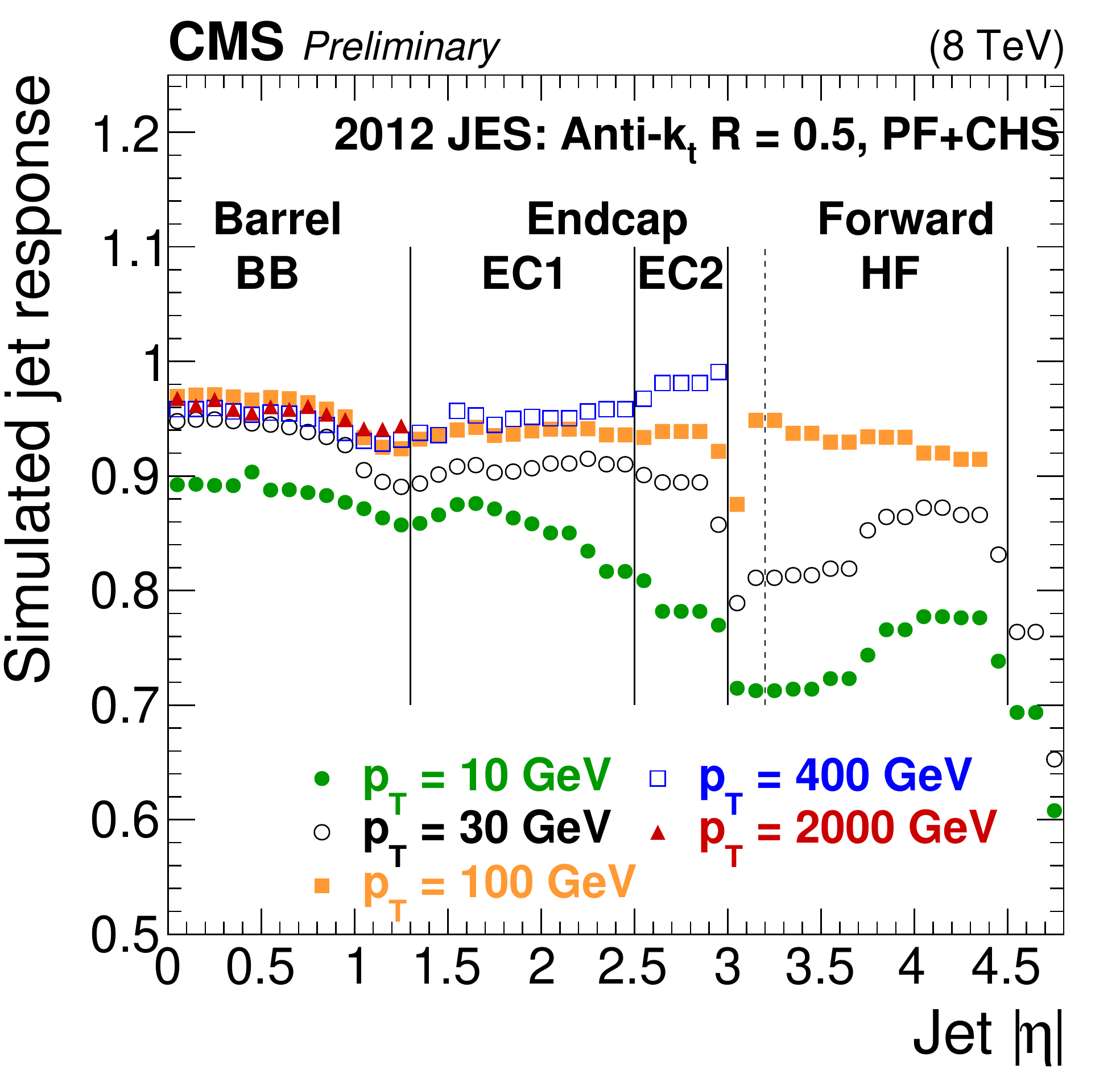} \label{JES_CMS}}
\caption{Average jet energy ($p_T$) response as a function of $\eta$ in different jet energy ($p_T$) bins for ATLAS \subref{JES_ATLAS} (CMS) \subref{JES_CMS}. Indicated in lines in \subref{JES_CMS} are the different detector boundaries. Figures taken from Ref. \protect\refcite{ref:ATLAScombo} and \protect\refcite{ref:CMScombo}.}
\label{fig:MCJES}
\end{figure}

The jet energy scale calibration is derived as a multiplicative factor that relates the reconstructed jet energy (or $p_T$), after pileup subtraction, to the truth-particle jet energy. The JES correction is obtained in an inclusive MC sample of dijet events and is defined as the inverse of the jet response (Eq.~\ref{RESPONSE}). In the case of ATLAS, the JES is obtained for the jet energy, whereas in CMS it is obtained for the jet $p_T$. The JES has a strong $\eta$ dependence, reflecting the different detector technologies and boundaries between the different sub-detectors. Figure~\ref{fig:MCJES} shows the $\eta$-dependence of the jet response in different jet energy bins for both ATLAS \cite{ref:ATLAScombo} and CMS \cite{ref:CMScombo}. The application of the JES equalizes these large differences in the jet response achieving a constant response as a function of $\eta$ and $p_T$ within $2\%$. 

%%------------------------------    

Since the jet energy calibration is derived in inclusive dijet events, it is only exact for the particular mixture of quark and gluon jets in this sample. However, the jet response for individual quark or gluon initiated jets is different due to the differences in jet fragmentation and particle composition and multiplicity~\cite{jespaper2011, CMSsubtraction}. Gluons tend to fragment into a  larger number of soft particles compared to quark jets, leading to a lower response on average since the calorimeter response is lower at low $p_T$. This effect is known as {\it flavor-dependence} of the jet response and is a major source of the jet energy scale systematic uncertainty. 

\subsubsection{Global sequential calibration}

As indicated in Fig.~\ref{fig:JESSCHEMES}, ATLAS applies a track-based post-calibration correction that attempts to reduce the differences in response between quark and gluon jets and also improves the jet energy resolution by correcting for differences in jet response as a function of jet-level (global) observables sensitive to the jet flavor. The corrections are applied sequentially, such that the average jet energy scale remains unchanged~\cite{refGSC}. In the case of jets constructed from calibrated topoclusters (LCW jets), two sequential corrections are applied, one based on the number of tracks above $p_T>1$~GeV associated to the jet, and another based on the jet width, defined as the average $\Delta R = \sqrt{(\Delta\eta)^2 + (\Delta\phi)^2}$ distance between the tracks associated to the jet and the calorimeter jet axis, weighted by the track $p_T$. A third correction is applied to high $p_T$ jets to account for energy leakage outside the calorimeter. This correction depends on the energy measured by the muon detector and matched to the jet.  More details about the global sequential calibration are found in Ref.~\refcite{refGSC}.

%---------------------------------------------------
\subsection{Residual in situ jet energy calibration}
%---------------------------------------------------

The residual in situ calibration exploits the balance of physics objects in the transverse plane to bring the jet response in data and MC to agreement. This correction is only applied to jets in data and is computed as:
\begin{equation}
JES^{~in~situ} =   \frac{ \langle p^{jet}_T/p^{ref}_T \rangle_{MC}}{\langle p^{jet}_T/p^{ref}_T \rangle_{data}}
\end{equation}
where the reference objects are photons, \Z~bosons, or other jets. 
%The importance of the in situ correction is that the uncertainty on the jet energy scale is given by the accuracy on the in situ jet response measurement rather than by the differences between data and MC or simulation based uncertainties. At the LHC, the large samples of dijet, $\gamma+$~jet, and Ä+jet events allow for a very precise determination of the jet energy response, leading to very small jet energy scale systematic uncertainties.  
The in situ calibration consists of three different corrections. First, dijet events are used to derive a relative $\eta$ inter-calibration where the response of forward jets are calibrated to the response of jets in the central region. The purpose of this correction is to equalize the jet response in $\eta$ after the MC jet energy scale correction. Next, an absolute $p_T$ calibration is derived for central jets using the balance of $\gamma$ and \Z~bosons recoiling against jets. The final step calibrates high $p_T$ jets, outside the acceptance of \gj, and \Zjets~events, relative to a system of recoiling, well calibrated, low $p_T$ jets.  

\subsubsection{Relative $\eta$-dependent calibration using dijet events}

Two main experimental techniques are used to measure the relative jet $p_T$ response as a function of $\eta$ in dijet events: the dijet balance and the missing $p_T$ fraction (MPF) methods. 

In the dijet balance method, the $p_T$ of a reference jet ($p^{ref}_T$) in the central region of the detector ($\eta<0.8$ in ATLAS, and $\eta<1.3$ in CMS) is balanced against a probe jet in the forward region ($p^{probe}_T$) exploiting the fact that, at leading order in QCD, these two jets should have the same $p_T$ and any imbalance in transverse momentum would be indicative of differences in detector response in different $\eta$ regions. The central region is chosen as  reference because of the uniformity of the jet response in this region and because it provides the best region to calibrate the absolute jet response using \gj, and \Zjets~events. The $p_T$ balance is defined by the asymmetry:
\begin{equation}
A = \frac{p^{probe}_T - p^{ref}_T}{p^{ave}_T}
\end{equation}
where $p^{ave}_T = (p^{probe}_T + p^{ref}_T)/2$ is the average $p_T$ of the two jets. The $\eta$ relative calibration factor is then defined as:
\begin{equation}
\frac{p^{probe}_T}{p^{ref}_T} = \frac{2+\avg A}{2- \avg A}
\end{equation}
where $\avg A$ is the mean of the asymmetry distribution in bins of $p^{ave}_T$ and $\eta$. ATLAS also uses a variation of this technique (the matrix method) where multiple reference regions are used to measure the jet response. 
%In the matrix method, the jet response of a given $\eta$ region is measured relative to all the reference regions and the $\eta$ calibration is obtained by solving a system of linear equations. 
The advantage of this technique is that it reduces the statistical uncertainty of the method, as many more jets are used. The matrix method is the primary technique used in ATLAS, with the dijet balance used as a validation. 

There are two main biases affecting in situ measurements of the jet response: the resolution bias, and the radiation bias. The resolution bias arises from the fact that both the reference and the probe objects have different energy resolution (depending on jet $\eta$) and that the jet $p_T$ spectrum is steeply falling. This means that any given bin of reconstructed jet $p_T$ will be sensitive to jets whose true $p_T$ fluctuated from nearby bins. The use of $p^{ave}_T$ in the definition of the relative calibration cancels out, on average, fluctuations on the $p_T$ due to resolution effects. The radiation bias is due to the $p_T$ imbalance caused by gluon radiation in the event. Two techniques are used to mitigate or correct for the effects of gluon radiation. ATLAS uses very tight event selection requirements on the azimuthal angle $\Delta \phi(jet_1, jet_2)$ between the two leading jets and the $p_T$ requirement on additional jets to enhance the 2 $\rightarrow$ 2 topology. CMS uses a soft radiation correction method that consists of measuring the jet response in bins of the $p_T$ fraction of the third leading jet ($\alpha = p^3_T/p^{ave}_T$) and linearly extrapolating the measured response in events with $\alpha=0.2$ to the limit $\alpha \rightarrow 0$.

The MPF technique uses the total hadronic recoil as a reference object to estimate the calorimeter response to jets. It is based on the fact that, at parton level, the reference object and the total hadronic recoil are perfectly balanced in the transverse plane:
\begin{equation}
\vec{p}^{~ref}_T + \vec{p}^{~recoil}_T = 0
\end{equation}
In the case of reconstructed objects, the above equation can be re-written as:
\begin{equation}
R_{ref}\vec{p}^{~ref}_T + R_{recoil}\vec{p}^{~recoil}_T = - \vec{\met}
\end{equation}
where $R_{ref}$ and $R_{recoil}$ are the detector responses to the reference object and the hadronic recoil respectively. 
In the limit of no activity outside the probe jet ($\alpha = 0$) $R_{recoil} = R_{probe}$, and assuming $R_{ref}=1$, the MPF response is defined as:
\begin{equation}
R_{MPF} = 1 + \frac{\vec{\met} \vec{p}^{~ref}_T}{(\vec{p}^{~ref})^2}
\end{equation}
The MPF method is sensitive to soft radiation only through differences in the response of the leading jet and the additional jets, making this technique less sensitive to gluon radiation than the dijet balance method. The residual bias to soft radiation is corrected similarly to the dijet balance method, by extrapolating the measured response at $\alpha=0.2$ to the limit $\alpha \rightarrow 0$.

\subsubsection{Absolute jet energy calibration using \gj~ and \Zjets~ events}

The in situ absolute jet energy scale is determined for central jets in the range 20~$\leq p_T \leq 800$~GeV exploiting the $p_T$ balance between a jet and a photon or Z boson (reference object). In CMS, both direct $p_T$ balance and MPF methods are used and combined, to reduce the uncertainty of the measurement. ATLAS uses the direct $p_T$ balance as primary method. In contrast to the $\eta$ relative calibration method, the jet energy response is measured in bins of the reference object rather than in bins of the average $p_T$. This is because the $p_T$ of the reference object is measured with much higher accuracy than the jet $p_T$, limiting the resolution bias. The effects of soft radiation are accounted for in the same way as for the relative $\eta$ calibration, extrapolating the response measurement to the limit of no additional jet activity in the event ($\alpha \rightarrow 0$). The advantage of \Zjets~ events is that it enables the response measurement down to very low $p_T$, which is very difficult to achieve in \gj~ events due to the higher trigger photon $p_T$ threshold and larger background contamination. On the other hand, \gj~ events provide a higher statistics sample to probe jets above 300~GeV. 

\subsubsection{Absolute jet energy calibration using multijet events}

In order to extend the absolute jet energy calibration to high jet $p_T$, the multijet balance method introduced in Ref.~\refcite{ref:MJB} is used. This technique exploits the balance of a high $p_T$ jet recoiling against a system of well calibrated low $p_T$ jets. CMS also uses the MPF method in multijet events, with the recoil system as the reference object. This technique covers the jet energy range $300 \leq p_T \leq 2$~TeV.

\subsubsection{Combination of in situ measurements}

\begin{figure}[h]
\centering
  \subfigure[]{\includegraphics[width=6.5cm]{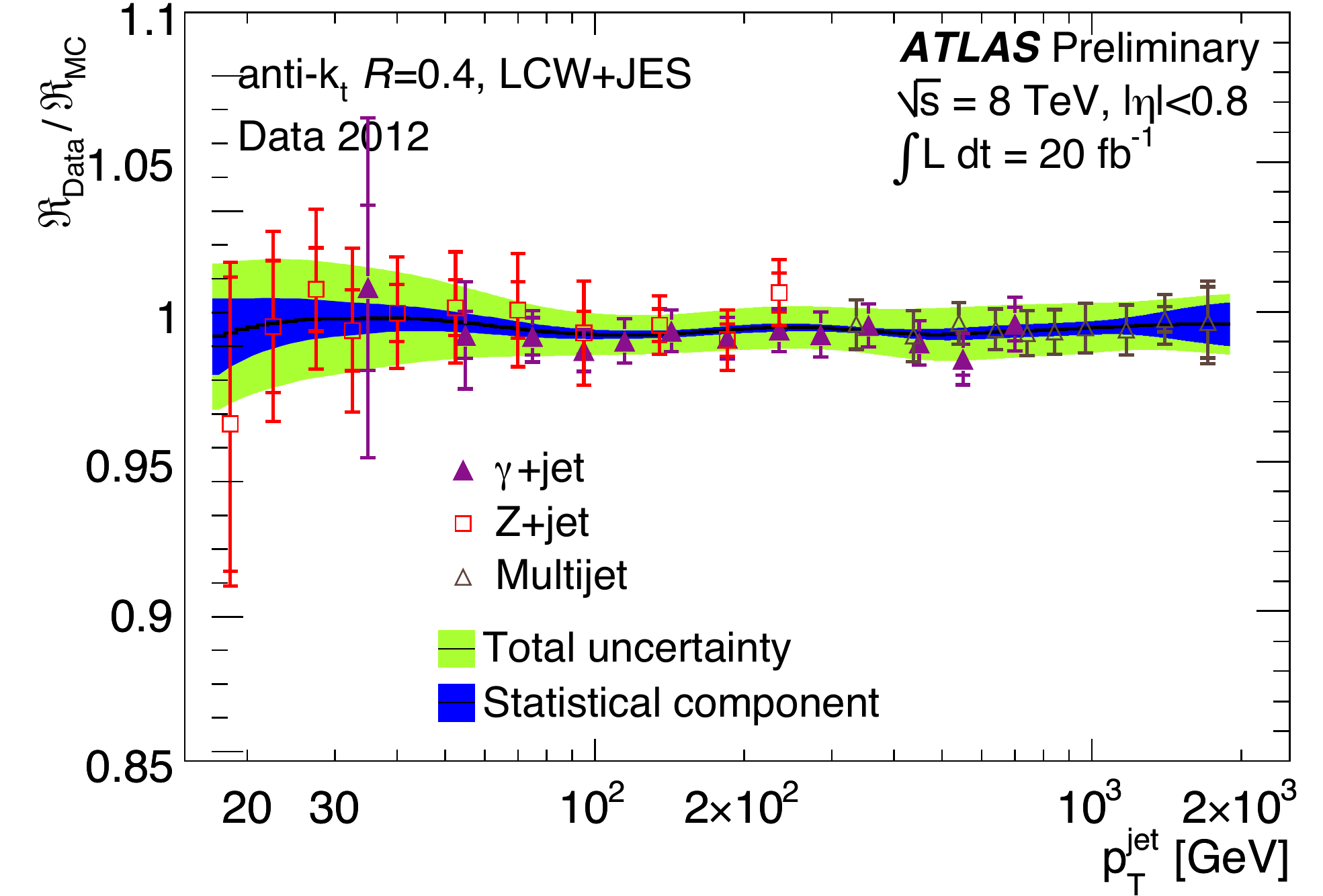} \label{fig:comboATLAS}}
  \subfigure[]{\includegraphics[width=5.0cm]{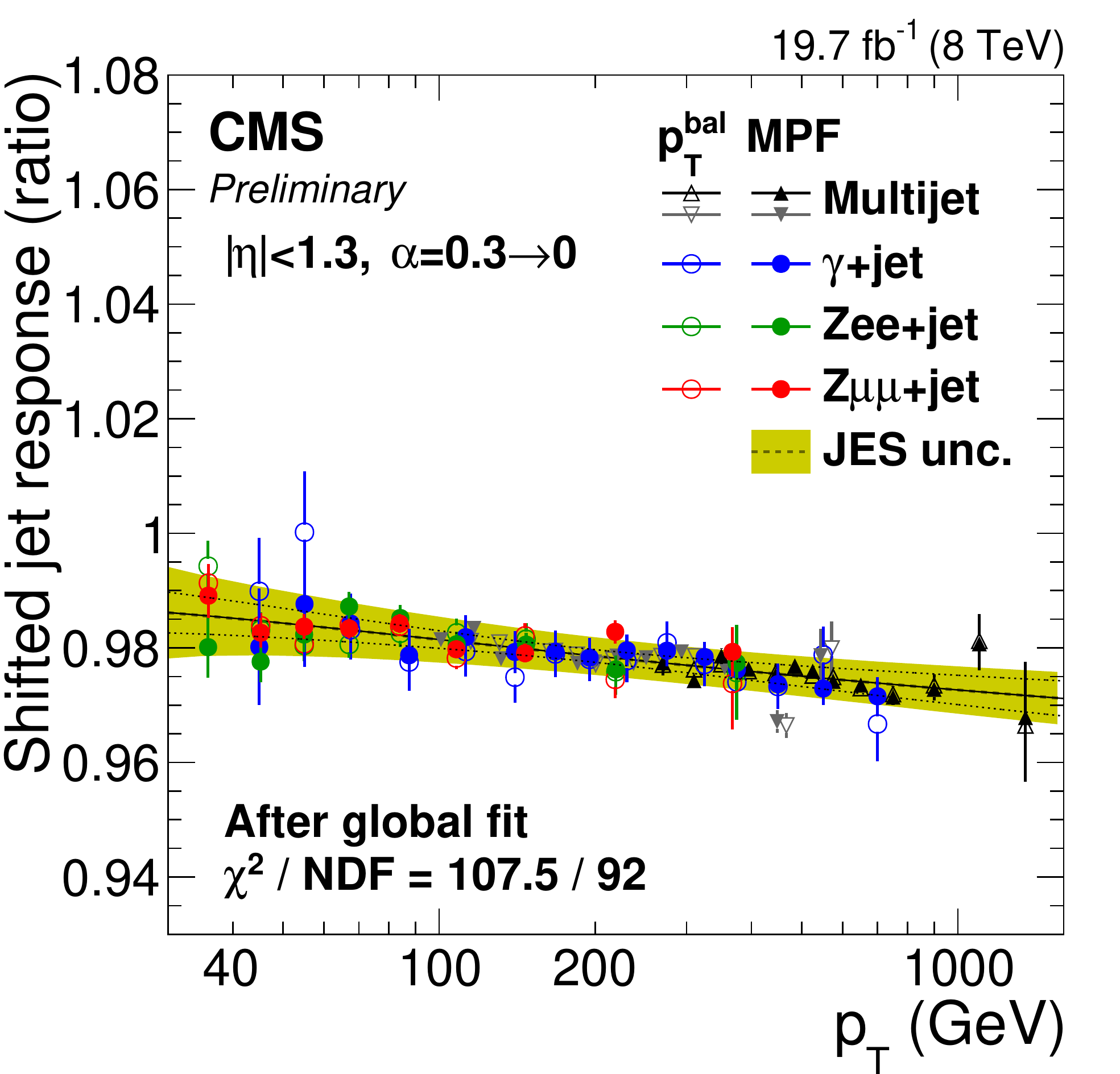} \label{fig:comboCMS}}
\caption{\subref{fig:comboATLAS} Ratio of jet response measured in data to jet response measured in MC using \gj, \Zjets, and multijet balance methods in ATLAS. Also shown are the combined correction (black line) and its uncertainty (green). \subref{fig:comboCMS} Ratio of jet response measured in data to jet response measured in MC using the $p_T$ balance and MPF methods in \gj, \Zjets, and multijet events. Also shown are the combined correction (black line) and its uncertainty (yellow). Figures taken from Ref. \protect\refcite{ref:ATLAScombo} and \protect\refcite{ref:CMScombo}.}
\label{fig:combo}
\end{figure}

The measurements of the jet response in data and MC from \Zjets, \gj, and multijet events are combined using the procedures described in Ref.~\refcite{ref:ATLAScombo} and 
Ref.~\refcite{ref:CMScombo}. The individual measurement of the absolute jet response data-to-MC ratio in each sample and its combination is shown in Fig.~\ref{fig:combo}. The three different measurements of jet response are in very good agreement. The difference from unity in the combined jet response shown in Fig.~\ref{fig:combo} defines the in situ jet energy scale calibration applied to jets in data. In CMS, the global fit of the residual absolute correction has two parameters, one for an absolute energy scale shift between data and MC, and another for the $p_T$ dependence under the assumption that the shape is given by a constant shift in the single pion response in the HCAL. In ATLAS, the data-to-MC response ratio is computed in fine $p_T$ bins for each in situ method using a second order spline interpolation method and the combined response is obtained as a weighted average of the interpolated contribution from each method.  The influence of the statistical fluctuations is reduced by applying a smoothing algorithm using a sliding Gaussian kernel. The systematic uncertainties are derived from the uncertainties on each of the in situ methods as described in section \ref{sec:uncertainty}.

\subsection{Jet energy scale systematic uncertainties}
\label{sec:uncertainty}

The total jet energy scale systematic uncertainty has many different sources associated to each of the jet energy corrections: pileup subtraction, $\eta$ relative calibration, and in situ absolute energy scale. In the case of in situ corrections, systematic uncertainties are primarily derived from changes in the data-to-MC response ratio under variations in the event selection. 
%This approach allows to disentangle physics modeling from detector effects, since variations in the selection criteria can suppress or enhance physics effects. 
Important contributions to the absolute jet energy scale uncertainty are the single particle response and fragmentation uncertainty. The former is estimated from the variations in the single particle response measured in test beam and collision data, and the latter is estimated from differences between the \Pythia and \Herwig event generators. Other sources of systematic uncertainties are due to the selection, calibration, and modeling of the reference objects.

The primary sources of systematic uncertainties for the $\eta$ relative calibration are due to the MC modeling of jets in the forward region, the modeling of the jet energy resolution, and the soft radiation bias correction. The MC modeling and soft radiation uncertainties are evaluated from differences in MC studies between the \Pythia and \Herwig event generators. The absolute energy scale uncertainties for the \gj~ and \Zjets~ methods are dominated by the lepton energy scale, the photon purity at low jet $p_T$, and the effect of particles outside the jet and particles from soft interactions not contributing to the $p_T$ balance (out-of-cone effects). Other sources of uncertainties include pileup, soft radiation correction, and jet energy resolution.  The uncertainties on the multijet balance method are dominated by the systematic uncertainties of the recoil system, MC modeling, and event selection cuts.  

\begin{figure}[h]
\centering
  \subfigure[]{\includegraphics[width=6.0cm]{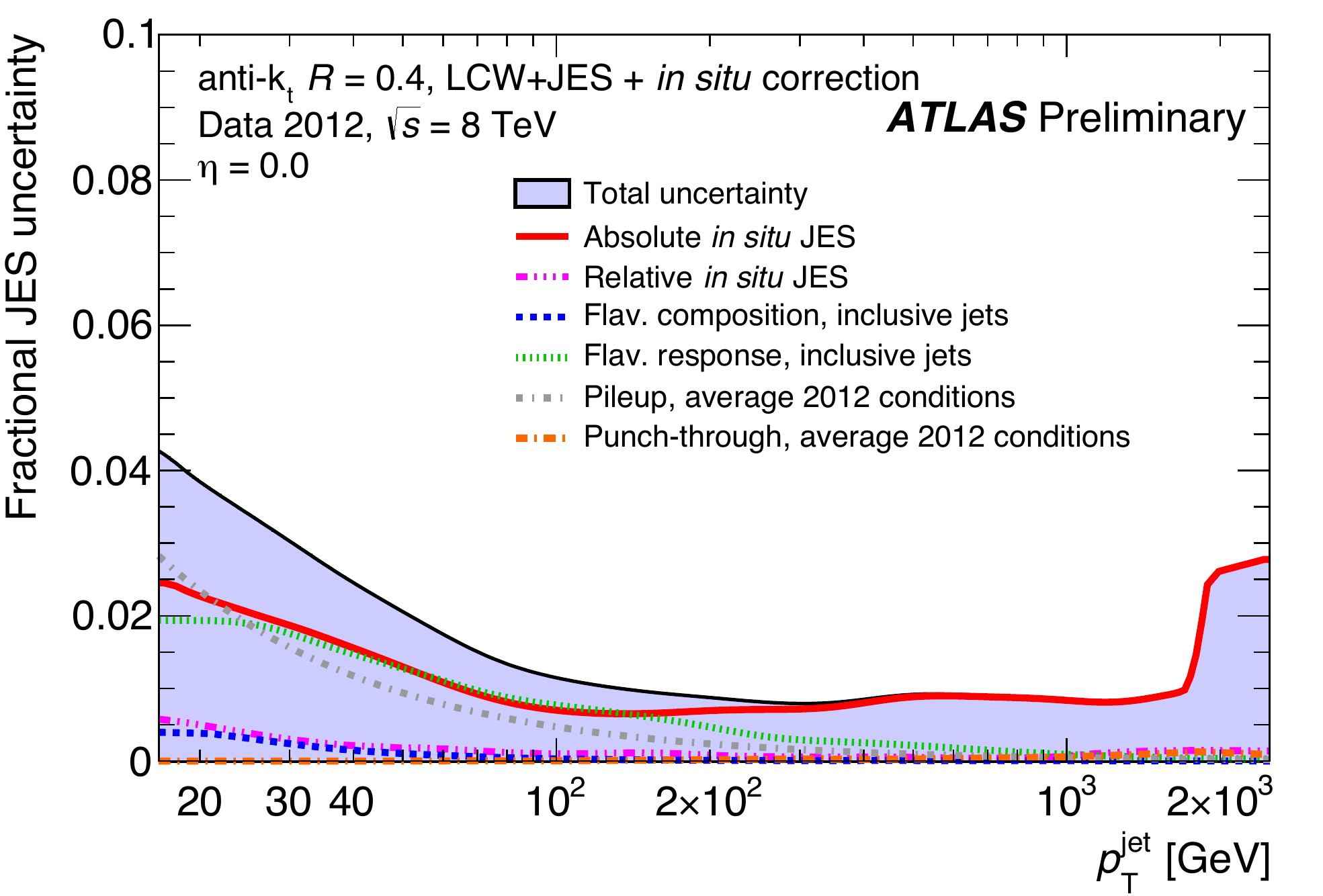} \label{fig:pt}}
  \subfigure[]{\includegraphics[width=6.0cm]{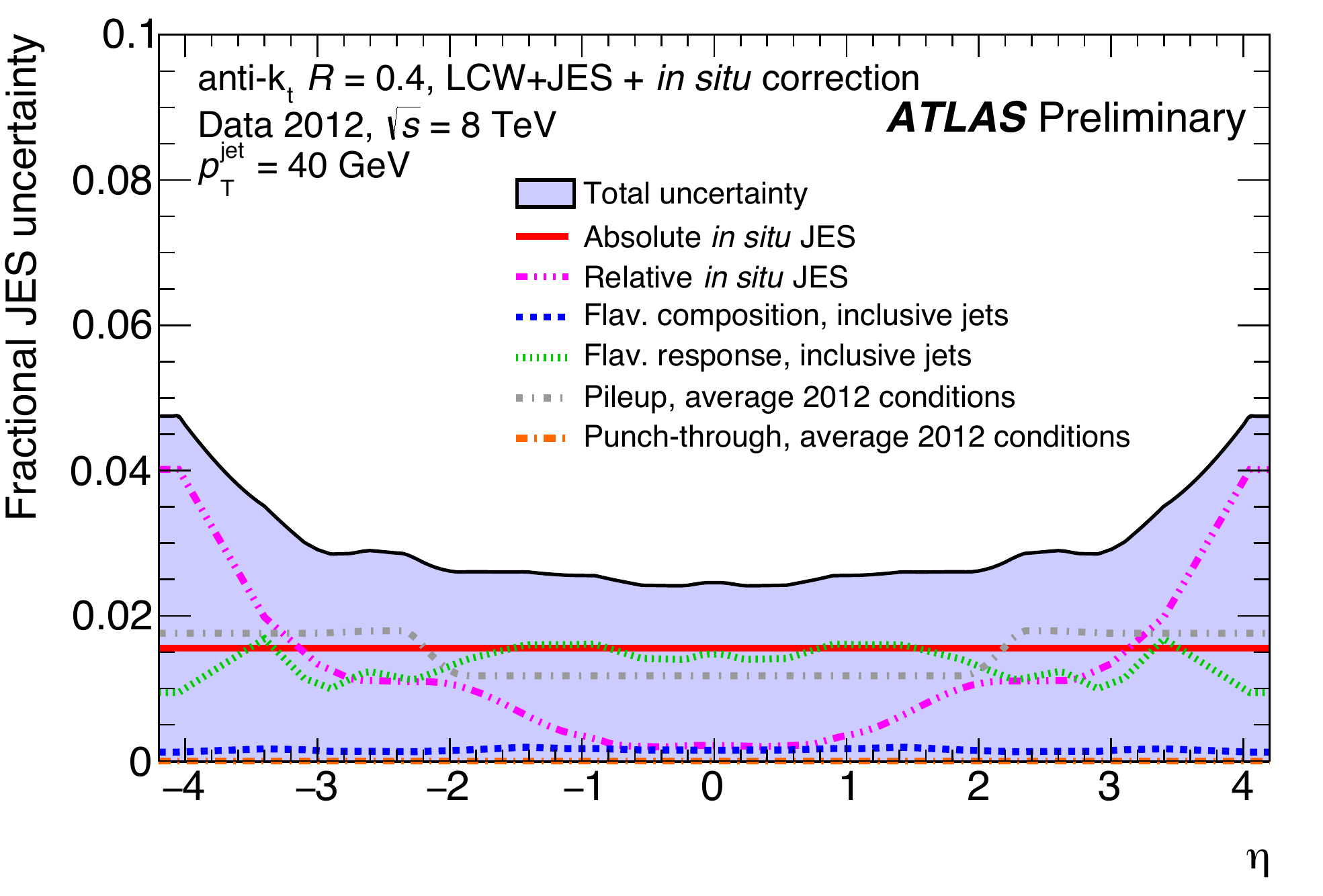} \label{fig:eta}}
\caption{ATLAS fractional jet energy scale systematic uncertainty as a function of $p_T$ at $|\eta| = 0.0$ \subref{fig:pt} and as function of $\eta$ for $p_T$=40GeV \subref{fig:eta}. The total uncertainty (all components summed in quadrature) is shown as a filled blue region topped by a solid black line. Average 2012 pileup conditions were used, and topology dependent components were taken from inclusive dijet samples. Figures taken from Ref. \protect\refcite{ref:ATLAScombo}.}
\label{fig:totalJESATLAS}
\end{figure}

\begin{figure}[h]
\centering
  \subfigure[]{\includegraphics[width=6.0cm]{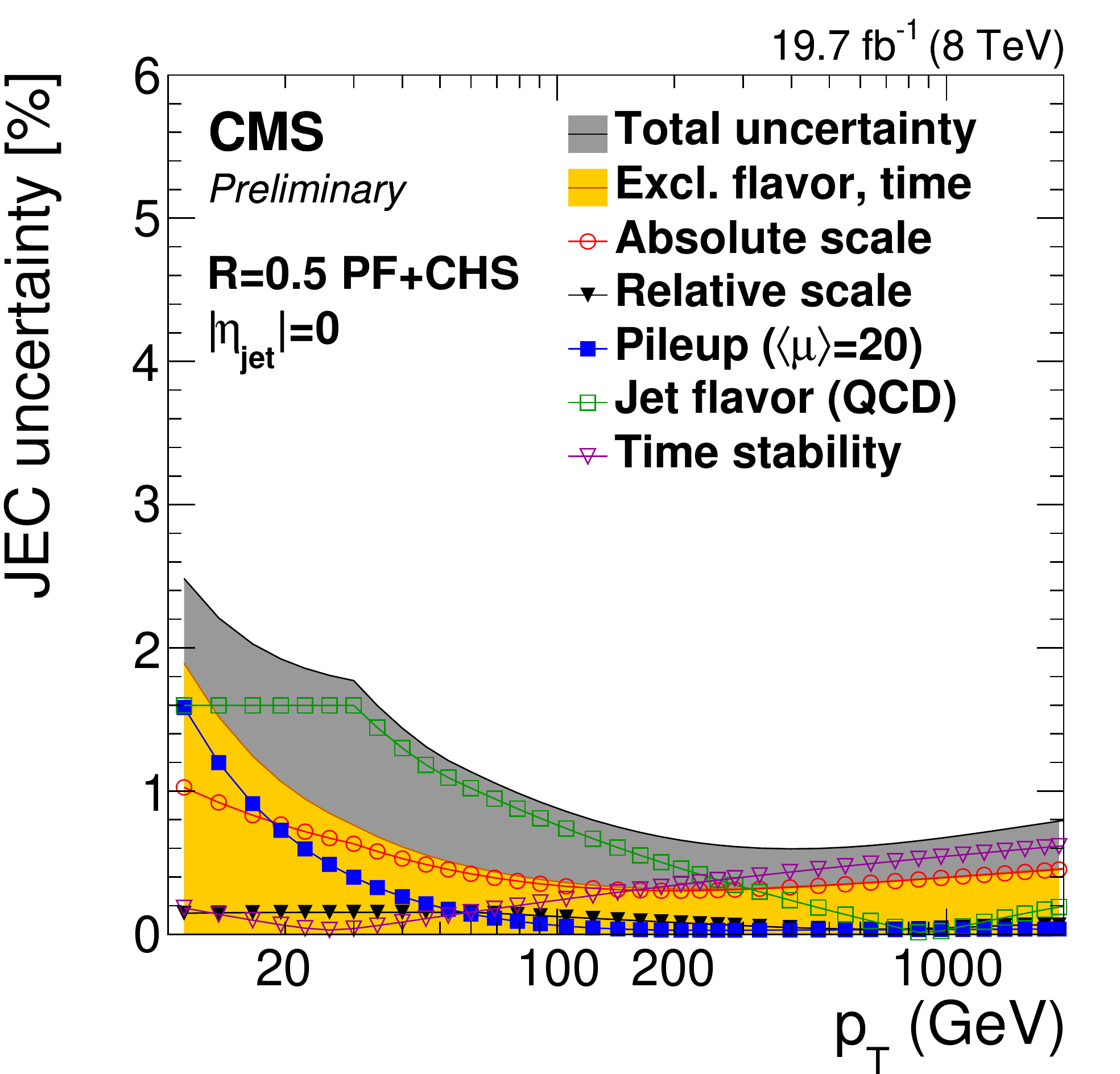} \label{fig:pt2}}
  \subfigure[]{\includegraphics[width=6.0cm]{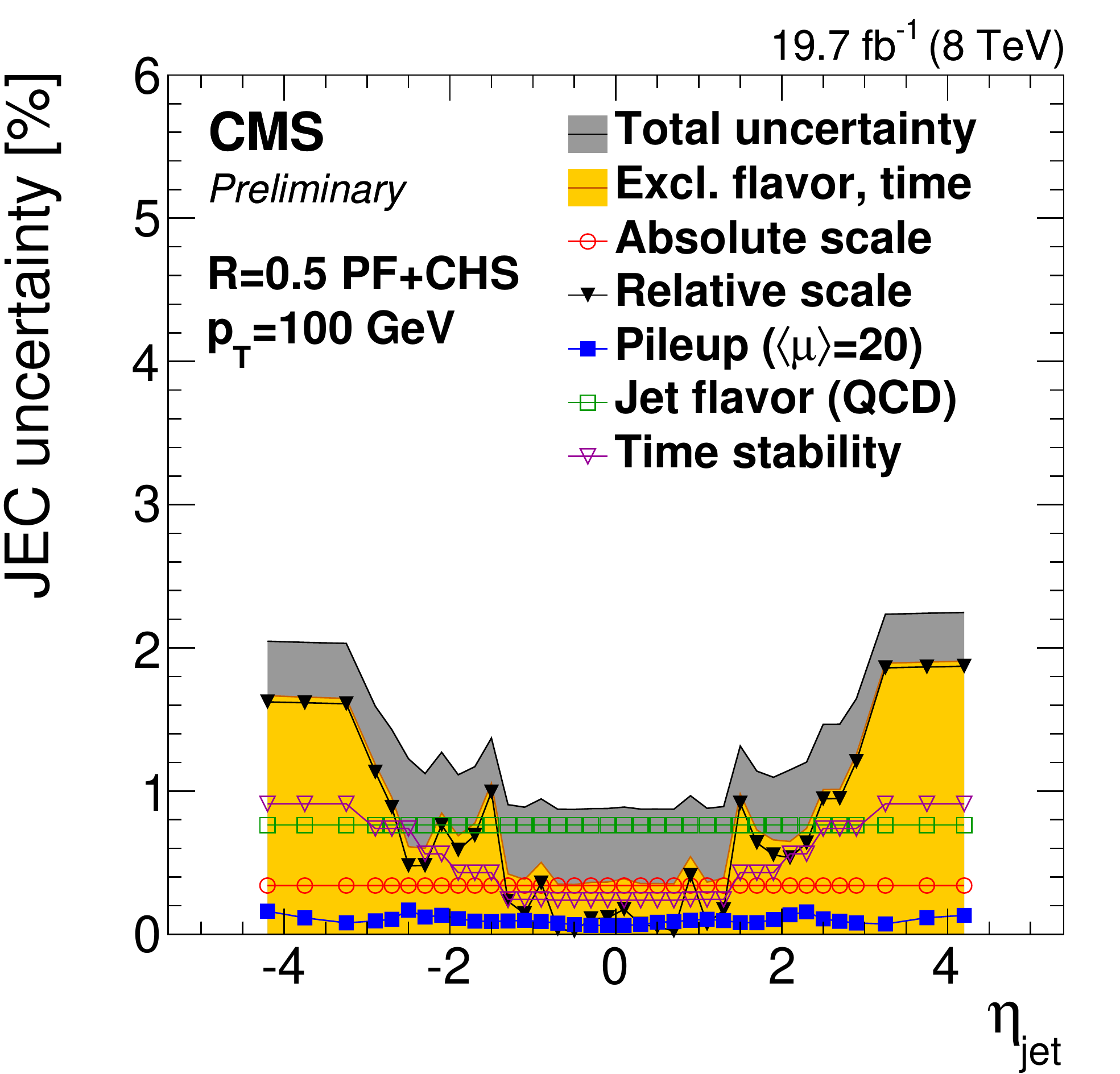} \label{fig:eta2}}
\caption{CMS fractional jet energy scale systematic uncertainty as a function of $p_T$ at $|\eta| = 0.0$ \subref{fig:pt2} and as function of $\eta$ for $p_T$=100GeV \subref{fig:eta2}. The total uncertainty (all components summed in quadrature) is shown as a filled grey region topped by a solid black line. Average 2012 pileup conditions were used, and flavor dependent components were taken from inclusive dijet samples. Figures taken from Ref. \protect\refcite{ref:CMScombo} }
\label{fig:totalJESCMS}
\end{figure}

Figures~\ref{fig:totalJESATLAS} and~\ref{fig:totalJESCMS} show the total jet energy scale uncertainty in ATLAS \cite{ref:ATLAScombo} and CMS \cite{ref:CMScombo} as a function of jet $p_T$ for central jets, and as a function of $\eta$ for low $p_T$ jets. The JES uncertainty is about 3-3.5\% at $p_T=20$~GeV, dominated by pileup uncertainties, and decreases to below 1-1.5\% for $100 \leq p_T \leq 2000$~GeV. 

\subsubsection{Jet flavor uncertainties}
\label{sec:flavor}

The energy response of jets originating from light quarks is different than for those originating from gluons\footnote{The flavor of of jet is defined as the flavor of the highest energetic matrix element parton within a cone of $\Delta R$ around the jet axis. This is a leading order definition and it is well defined for the \Pythia and \Herwig generators.}, as shown in Ref.~\refcite{jespaper2011} and Ref.~\refcite{CMSsubtraction}. This is due to differences in fragmentation between the two types of jets. For the same truth-particle jet $p_T$, jets originating from gluons tend to have more particles of softer $p_T$ than jets originating from light quarks. In addition, gluon initiated jets tend to have a wider angular energy profile.
The lower energy response of the calorimeter to low $p_T$ particles, combined with threshold effects related to the energy density distribution within jets, result in gluon initiated jets having a lower energy response than quark initiated jets. These differences in flavor response are significantly reduced, but not eliminated, with the use of particle flow in CMS and the GSC calibration method in ATLAS (not applied in Fig.~\ref{fig:totalJESATLAS}). Additionally, it is observed that the response of gluon jets varies between different MC generators due to differences in jet fragmentation. Since the jet  in situ calibration has been derived in samples with a specific flavor content, the application of the JES is only strictly correct when it is applied to samples with the same flavor composition. The application of the JES to physics samples with different flavor content therefore requires two additional flavor uncertainties, one to account for the differences in flavor composition between the calibration sample and physics sample (flavor composition) and another to account for the difference in the gluon jet response between different MC generators (flavor response). If the flavor composition of a physics sample is known, the flavor composition uncertainty can be reduced.

\subsection{b-jet energy calibration}

\begin{figure}[h]
\centering
  \subfigure[]{\includegraphics[width=5.3cm]{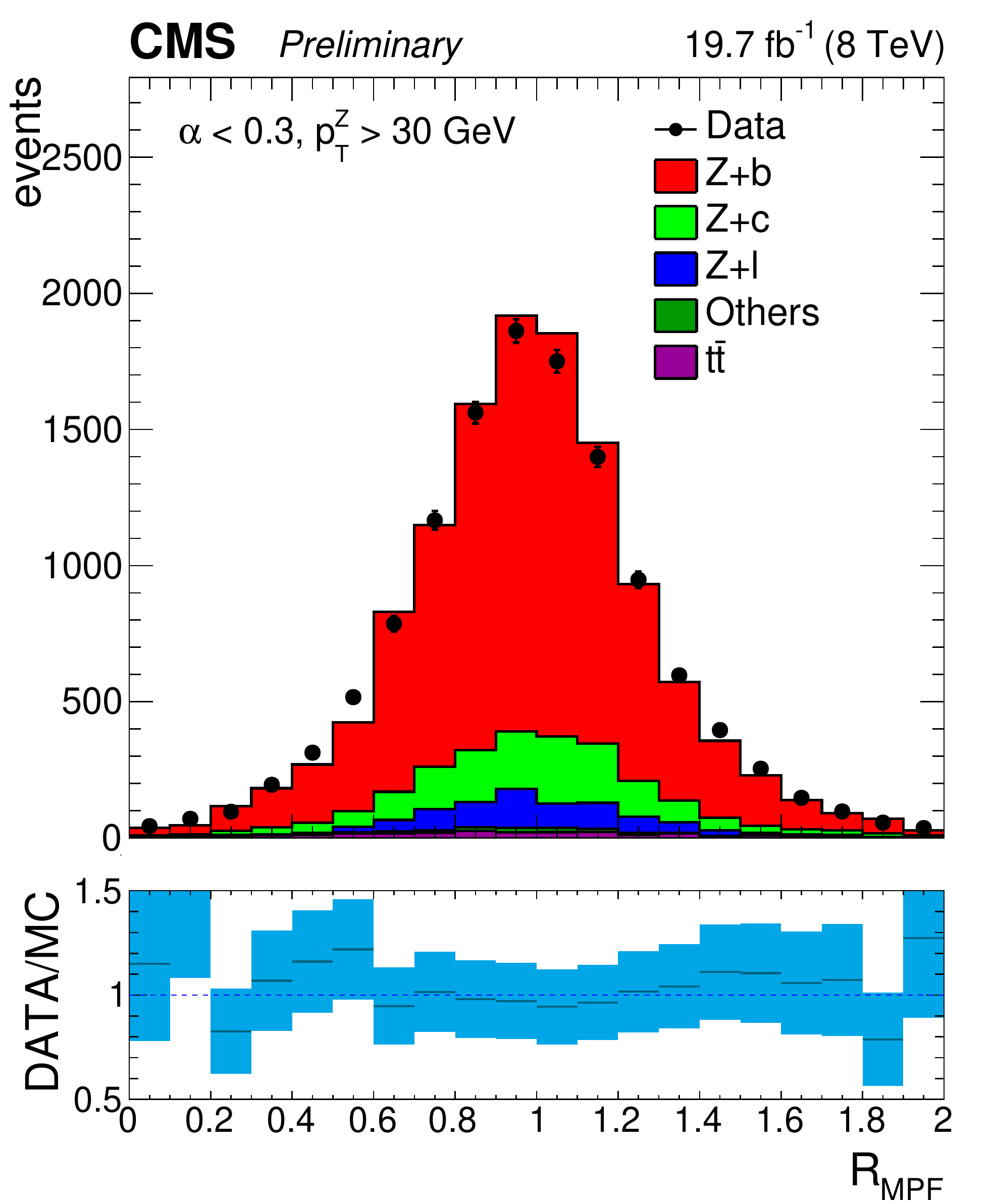} \label{BJET1}}
  \subfigure[]{\includegraphics[width=6.8cm]{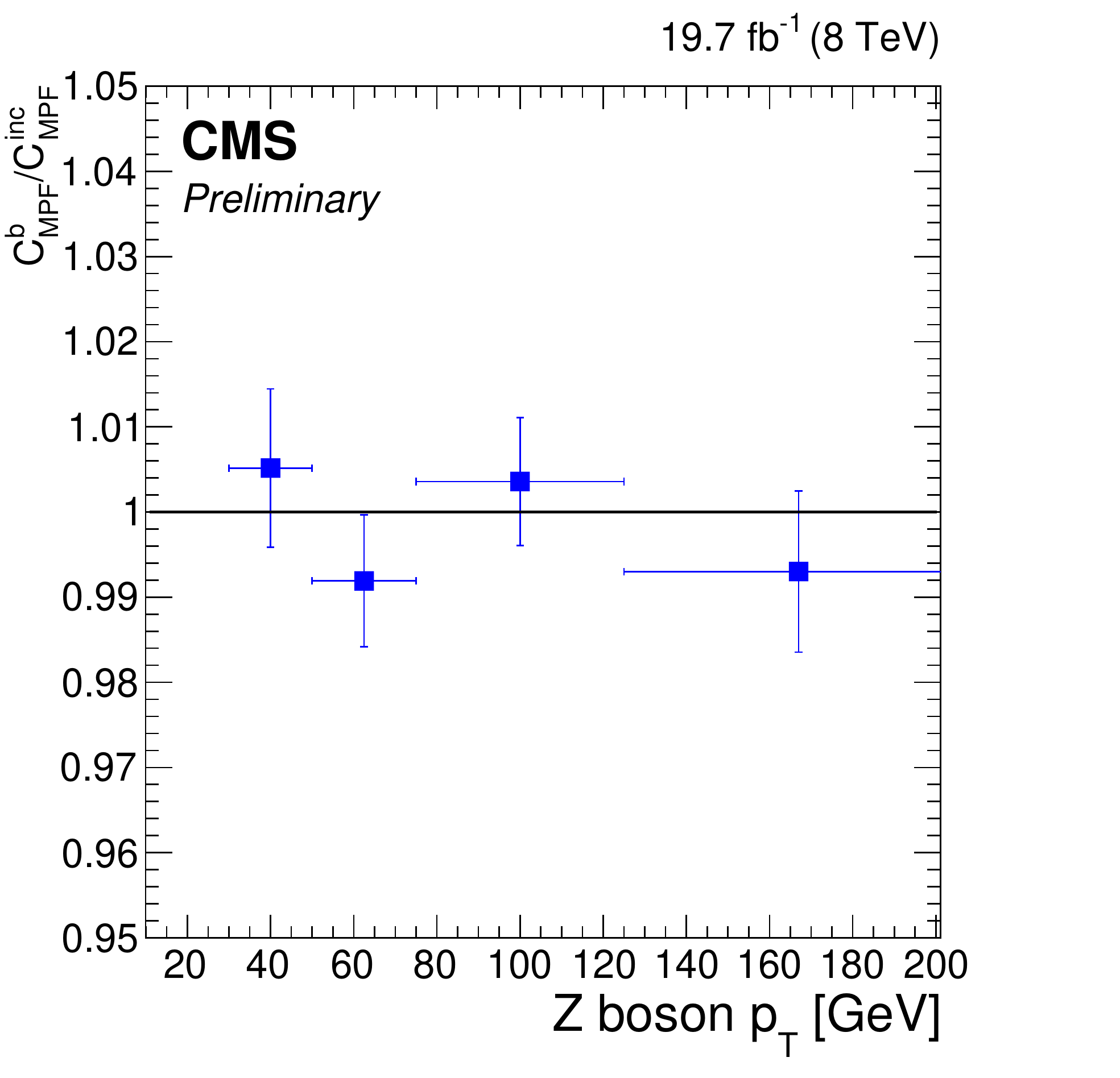} \label{BJET2}}
\caption{  \subref{BJET1} MPF response in $Z$+$b$-jet events in data and Monte Carlo. \subref{BJET2} Double ratio of the data-to-MC jet response in $Z$+$b$ events to that of inclusive $Z$+jet events as a function of the $Z$ boson momentum. Figures taken from Ref. \protect\refcite{CMSb}.}
\label{fig:bjet}
\end{figure}

Both ATLAS and CMS studied the jet energy scale uncertainty for $b$-jets. ATLAS used both a Monte Carlo based method, and an in situ approach, comparing the measured jet energy to that of independent, well calibrated, track-jets \cite{jespaper2011}. This ratio is evaluated for inclusive and $b$-tagged jets in dijet events. CMS utilized the \Zjets~ balance in Z+$b$ events \cite{CMSb}. The event selection requires the leading jet to be $b$-tagged and the $b$-jet response is computed using both the direct $p_T$ balance and MPF methods. This selection has a $70-80\%$ purity for Z+$b$-jets and the results are extrapolated to 100\% purity. Figure~\ref{BJET1} shows that the Z+$b$-jet response distribution with a $\alpha = p_{T,j3}/p_{T, ave} < 0.3$ for data and \PythiaS Monte Carlo are in very good agreement. A relative systematic uncertainty is computed as the double ratio of the residual data-to-MC jet response ratio in Z+$b$-jet events to that of inclusive \Zjets~ events. The measured response ratio is shown as a function of Z boson $p_T$ in Fig.~\ref{BJET2}. The inclusive response ratio is C = 0.998 $\pm$ 0.004 (stat.) $\pm$ 0.004 (syst.) which is consistent with unity and comparable with the $b$-jet flavor uncertainty of about $0.5\%$.

%-----------------------------------------------------------
\section{Conclusions}
%-----------------------------------------------------------

Jets are key physics signatures for the analysis of events at the LHC. They are fundamental to most precision measurements and searches for physics beyond the Standard Model at the LHC. At the same time, jet reconstruction and calibration face unprecedented challenges in the high pileup environment and the high center-of-mass energy of the LHC. Despite these challenges, ATLAS and CMS have developed  sophisticated techniques for the reconstruction and calibration of jets that resulted in an extraordinary level of accuracy in jet energy measurements. Such precision in the determination of the jet energy scale has broadly impacted the LHC physics program, enhancing the discovery potential to new physics and improving the precision of Standard Model measurements. Several factors contributed to this success. First, key contributions from the theory community and a tight collaboration between theory and experiments which led to the early adoptions of new techniques. Examples include the anti-$k_t$ algorithm and the jet areas pileup subtraction method. Second, the extensive use of tracking information and its combination with calorimeter signals to improve the precision of the jet energy measurements. ATLAS and CMS developed different approaches to combine tracking with calorimeter measurements, motivated by their different detector strengths. Finally, the optimal and full use of the detector capabilities in the definitions of the inputs to jet reconstruction. Topological clustering in ATLAS makes use of the full 3-dimensional capabilities of its finely segmented calorimeters to reconstruct and calibrate individual components of the jets, while CMS integrates tracking with calorimeter measurements in the elaborate and high performing particle flow candidates. This sophistication in the inputs to jet reconstruction allowed to mitigate many of the challenges for jet reconstruction, in particular pileup  (ATLAS topological clustering noise suppression, and  CMS charged hadron subtraction) and calorimeter response (ATLAS local cluster weighting, and CMS particle flow). This significantly reduced the reliance on simulations to determine the jet energy scale. Furthermore, large samples of \gj, \Zjets, and multijet events enabled extremely precise measurements of the jet energy response in the data that has led to a jet energy scale uncertainty of less than 1\% for jets in the central detector region with $100<p_T<2000$ GeV.

%-----------------------------------------------------------
\section*{Acknowledgments}
%-----------------------------------------------------------

I would like to thank Michael Begel, Philip Harris, Ia Iasvilli, David Lopez Mateos, and Mikko Voutilainen for their helpful comments and feedback on the manuscript. This work is supported by the US Department of Energy (DOE) Early Career Research Program and grant DE-AC02-76SF00515.

%-----------------------------------------------------------


\begin{thebibliography}{0}    %for 1 digit
%-----------------------------------------------------------

\bibitem{lumi} ATLAS Collaboration, https://twiki.cern.ch/twiki/bin/view/AtlasPublic/

LuminosityPublicResults.

\bibitem{ATLASdet} ATLAS Collaboration, {\it JINST} {\bf 3} S08003 (2008).

\bibitem{CMSdet} CMS Collaboration, {\it JINST} {\bf 3} S08004 (2008).

\bibitem{figATLAS} ATLAS Collaboration, {\it Eur.Phys.J.} {\bf C71} 1593 (2011). 

\bibitem{figCMS} T. Sakuma, T. McCauley, {\it arXiv:1311.4942 [physics.ins-det]}.

\bibitem{ref:pythia} T. Sjostrand, S. Mrenna, and P. Z. Skands, {\it Comput. Phys. Commun.} {\bf 178} 852Ð867 (2008).

\bibitem{ref:A2} ATLAS Collaboration, ATL-PHYS-PUB-2012-003 (2012).

\bibitem{ref:CT10} H.-L. Lai et al., {\it Phys. Rev. D} {\bf 82} 074024 (2010).

\bibitem{ref:herwig1} M. Bahr et al., Herwig++ Physics and Manual, {\it Eur. Phys. J.} {\bf C58} 639Ð707 (2008).

\bibitem{ref:herwig2} S. Gieseke et al., Herwig++ 2.5 Release Note, {\it arXiv:1102.1672 [hep-ph]}.

\bibitem{ref:datadriven} ATLAS Collaboration, ATLAS-CONF-2015-017 (2015).

\bibitem{ref:ATLAS19} ATLAS Collaboration, {\it Eur. Phys. J.} {\bf C70} 823Ð874 (2010).

\bibitem{ref:G4} GEANT4 Collaboration, {\it Nucl. Instrum. Meth. A} {\bf 506} 250Ð303 (2003).

\bibitem{ref:QGSP} G. Folger and J. Wellisch, {\it arXiv:0306007 [nucl-th]}.

\bibitem{ref:pythia6} T. Sjostrand, S. Mrenna, and P. Z. Skands, {\it JHEP} {\bf 05} 026 (2006).

\bibitem{refZ2} R. Field, {\it Acta Phys. Polon.} {\bf B42} 2631-2656 (2011).

\bibitem{ref:madgraph} J. Alwall et. al., {\it JHEP} {\bf 1106} 128 (2011). 

\bibitem{ref:powheg} S. Frixione, P. Nason, and C. Oleari, {\it JHEP} {\bf 0711} 070 (2007).

\bibitem{ref:topo} W. Lampl et al., ATL-LARG-PUB-2008-002 (2008).

\bibitem{LEP1} I.G. Knowles and G.D. Lafferty, {\it J. Phys. G} {\bf 23} 731 (1977).

\bibitem{refLEP} ALEPH Collaboration, {\it Nucl. Instrum. Meth. A} {\bf 360}, 481 (1995).

\bibitem{ref3} ATLAS Collaboration, {\it Eur. Phys. J. C} {\bf 73}, 2304 (2013).

\bibitem{ref62} C. Cojocaru {\it et al.} {\it Nucl. Instrum. Meth. A} {\bf 531}, 481-514 (2004).

\bibitem{ref66} E. Abat {\it et al.} {\it JINST} {\bf 5} P11006 (2010).

\bibitem{ref67} M. Aharrouche {\it et al.} {\it Nucl. Instrum. Meth. A} {\bf 614}, 400-432 (2010).

\bibitem{refpflow} CMS  Collaboration, PAS PFT-09-001 (2009).

\bibitem{refAKT} M. Cacciari, G. P. Salam, and G. Soyez, {\it JHEP} {\bf 04} 063 (2008).

\bibitem{refFASTJET} M. Cacciari, G. P. Salam, G. Soyez, fttp://fastjet.fr/.

\bibitem{refAKT2} M. Cacciari, J. Rojo, G. P. Salam, G. Soyez {\it Eur. Phys. J. C} {\bf 71}, 1539 (2011).

\bibitem{LAR1} N. Buchanan et al., {\it JINST} {\bf 3} P09003 (2008).

\bibitem{LAR2} H. Abreu et al., {\it JINST} {\bf 5} P09003 (2010).

\bibitem{refareas} M. Cacciari and G. P. Salam, {\it Phys. Lett. B} {\bf 659}, 119Ð126 (2008).

\bibitem{JVF} ATLAS Collaboration, ATLAS-CONF-2013-083 (2013).

\bibitem{ref:JVT} ATLAS Collaboration, ATLAS-CONF-2014-018 (2014).

\bibitem{PileupID} CMS Collaboration, JME-13-005 (2013).

\bibitem{jespaper2011} ATLAS Collaboration, {\it Eur. Phys. J. C}, 75:17 (2015).

\bibitem{fJVT} ATLAS Collaboration, ATL-PHYS-PUB-2015-034 (2015).

\bibitem{CMSsubtraction} CMS Collaboration, {\it JINST} {\bf 6} P11002 (2011).

\bibitem{CMSb} CMS Collaboration, JME-13-001 (2014)

%\bibitem{refQG} ATLAS Collaboration, {\it Eur. Phys. J. C} {\bf 74}, 3023 (2014).

\bibitem{refGSC} ATLAS Collaboration, ATLAS-CONF-2015-002, CERN (2015).

\bibitem{ref:MJB} ATLAS Collaboration, ATLAS-CONF-2013-003, CERN (2013).

\bibitem{ref:ATLAScombo} ATLAS Collaboration, ATLAS-CONF-2015-037, CERN (2015).

\bibitem{ref:CMScombo} CMS Collaboration, CMS DP-2015/044, CERN (2015).

\bibitem{ref:CMScones} CMS Collaboration, https://twiki.cern.ch/twiki/bin/view/CMSPublic/

 MultipleConeSizes14.


%\bibitem{}


\end{thebibliography}
\end{document}